\documentclass[11pt,letterpaper]{article}
\usepackage[margin=1in]{geometry}
\IfFileExists{lmodern.sty}{\usepackage[T1]{fontenc}\usepackage{lmodern}}{}
\usepackage{microtype}
\usepackage{amsmath,amssymb,amsthm,mathtools}
\usepackage{xcolor}
\usepackage{tikz}
\usetikzlibrary{arrows.meta,calc,positioning,decorations.pathreplacing}
\usepackage{enumitem}
\setlist{itemsep=2pt,topsep=4pt}
\setlist[itemize]{label=$\bullet$}
\usepackage{booktabs,array}
\definecolor{linkblue}{RGB}{24,63,130}
\definecolor{figblue}{RGB}{24,63,100}
\definecolor{figwarm}{RGB}{157,79,22}
\definecolor{figgreen}{RGB}{30,120,60}
\definecolor{figred}{RGB}{170,40,40}
\definecolor{figviolet}{RGB}{112,48,160}
\usepackage[colorlinks=true,linkcolor=linkblue,citecolor=linkblue,urlcolor=linkblue]{hyperref}
\usepackage[capitalize,nameinlink,noabbrev]{cleveref}
\def\pdfauthornames{Hani Ahmadzadeh and Morteza Alimi}
\hypersetup{pdftitle={An EPTAS for Max-Cut on Interval Graphs},pdfauthor={\pdfauthornames}}
\usepackage[textsize=tiny,textwidth=2cm,color=green!50!gray]{todonotes} 

\newif\iffinal
\finalfalse

\AtBeginDocument{\iffinal\ifx\pdfauthornames\empty\PackageError{authorfill}{Unfilled author item: PDF author metadata (\string\pdfauthornames)}{Fill every author item before the final build.}\fi\fi}

\usepackage{aliascnt}
\newtheorem{theorem}{Theorem}[section]
\newaliascnt{lemma}{theorem}
\newtheorem{lemma}[lemma]{Lemma}
\aliascntresetthe{lemma}
\newaliascnt{proposition}{theorem}
\newtheorem{proposition}[proposition]{Proposition}
\aliascntresetthe{proposition}
\crefname{lemma}{Lemma}{Lemmas}
\Crefname{lemma}{Lemma}{Lemmas}
\crefname{proposition}{Proposition}{Propositions}
\Crefname{proposition}{Proposition}{Propositions}
\newcommand{\restatedthmnumber}{}
\newtheorem*{restatedtheoreminner}{Theorem~\restatedthmnumber}
\newenvironment{restatedtheorem}[1]
  {\renewcommand{\restatedthmnumber}{\ref*{#1}}\begin{restatedtheoreminner}}
  {\end{restatedtheoreminner}}

\newcommand{\eps}{\varepsilon}
\newcommand{\agw}{\alpha_{\mathrm{GW}}}
\DeclareMathOperator{\cut}{cut}
\DeclareMathOperator{\mc}{mc}
\DeclareMathOperator{\mb}{mb}
\DeclareMathOperator{\cw}{cw}
\DeclareMathOperator{\anc}{anc}
\DeclareMathOperator{\depth}{depth}
\DeclareMathOperator{\val}{val}
\newcommand{\Loc}{\mathrm{Loc}}
\newcommand{\cA}{\mathcal{A}}
\newcommand{\cR}{\mathcal{R}}
\newcommand{\cO}{\mathcal{O}}
\newcommand{\cD}{\mathcal{D}}
\newcommand{\NP}{\mathrm{NP}}
\newcommand{\Pclass}{\mathrm{P}}

\title{An EPTAS for Max-Cut on Interval Graphs}
\author{%
  Hani Ahmadzadeh\\
  {\normalsize Faculty of Mathematics}\\
  {\normalsize K.~N.~Toosi University of Technology, Tehran, Iran}\\
  {\normalsize\texttt{ahmadzadeh@kntu.ac.ir}}%
  \and
  Morteza Alimi\\
  {\normalsize Institute of Computer Science}\\
  {\normalsize University of Augsburg, Germany}\\
  {\normalsize\texttt{morteza.alimi@uni-a.de}}%
}
\date{}

\begin{document}

\hypersetup{pageanchor=false}
\begin{titlepage}
\maketitle
\thispagestyle{empty}
\begin{abstract}
\normalsize\noindent Whether Max-Cut is NP-hard on interval graphs, the intersection graphs of intervals on a line, was a long-standing open question, answered positively by Adhikary, Bose, Mukherjee, and Roy [SoCG 2021]. The best approximation ratio previously known for this class, due to Ahn, DeHaan, Kim, and Lee [Approx 2025], exceeds the Goemans--Williamson ratio of about $0.878$ by $10^{-34}$. We give an efficient polynomial-time approximation scheme (EPTAS): for every $\eps\in(0,1)$, a cut with at least $(1-\eps)$ times the maximum number of cut edges can be computed deterministically in time $2^{O(\eps^{-2}\log^3(2/\eps))}\cdot n^2$ on an unweighted $n$-vertex interval graph. This is the strongest type of scheme possible unless $\mathrm{P}=\mathrm{NP}$, since a fully polynomial-time scheme would solve the problem exactly.
The same holds for Max-Bisection, where the two sides of the cut must have sizes $\lfloor n/2\rfloor$ and $\lceil n/2\rceil$.

The algorithm rests on a construction that also yields a structural theorem of independent interest: every interval graph with $m$ edges can be turned into an interval graph of clique-width $O(\eps^{-2}\log^2(2/\eps))$ by deleting at most $\eps m$ edges, a bound that remains meaningful for sparse graphs. The construction replaces the endpoints by their ranks and uses a suitably shifted hierarchy of cells obtained by repeated halving. Intervals meeting cell boundaries at their own scale have small total degree, and their vertices are made isolated; each remaining interval is rounded to a grid adapted to its own length and stored in the hierarchy, so that the effect of long intervals on much smaller subproblems is captured by counts. Combined with Lampis's randomized scheme for graphs of bounded clique-width, the structural theorem already gives a randomized EPTAS for Max-Cut; our algorithm is instead a deterministic dynamic program over the hierarchy, which also handles Max-Bisection.

In contrast, we show that for every $\eps>0$ it is NP-hard to approximate Max-Cut within $84/85+\eps$ on chordal graphs,  which contain all interval graphs. Previously, approximation schemes on chordal graphs had been excluded only under the Small Set Expansion Hypothesis and with randomized reductions. 
\end{abstract}
\end{titlepage}

\pagenumbering{gobble}
\setcounter{tocdepth}{2}
\clearpage
\hypersetup{pageanchor=true}
\pagenumbering{arabic}

\section{Introduction}\label{sec:intro}

In the Max-Cut problem we are given a graph $G=(V,E)$ and ask for a set $S\subseteq V$ that maximizes the number $\cut_G(S)$ of edges with exactly one endpoint in $S$; we write $\mc(G)$ for this maximum, and we say that an algorithm approximates Max-Cut within a factor $\beta\le1$ if it always returns a set $S$ with $\cut_G(S)\ge\beta\cdot\mc(G)$. The randomized polynomial-time algorithm of Goemans and Williamson~\cite{goemans1995}, based on semidefinite programming, achieves the factor $\agw\approx0.878$ in expectation. Approximating Max-Cut within $\agw+\eps$ was shown to be NP-hard under the Unique Games Conjecture, for every $\eps>0$~\cite{kkmo2007,moo2010}.\footnote{\label{fn:ugc-update}In October 2026, OpenAI released the AI-generated manuscript \emph{The Unique Games Theorem}, with a Lean formalization~\cite{openai2026ugc}. Its stated result would remove the UGC assumption from this bound. } 
The classical $16/17+\eps$ hardness bound, for every $\eps>0$, does not use UGC: it follows from the results of H{\aa}stad~\cite{hastad2001} and of Trevisan, Sorkin, Sudan, and Williamson~\cite{tssw2000}, extended to unweighted graphs by Crescenzi, Silvestri, and Trevisan~\cite{cst2001}. On restricted graph classes the problem can be much easier: it is solvable in polynomial time on planar graphs~\cite{hadlock1975} and on graphs of bounded treewidth~\cite{bj2000}, and it admits approximation schemes on dense graphs~\cite{akk1999,fdlv1996}; Jansen, Karpinski, Lingas, and Seidel gave approximation schemes for the closely related Max-Bisection problem on planar graphs and on unit disk graphs given with a geometric representation~\cite{jkls2005}. For many natural classes, however, the approximability of Max-Cut has remained unclear. In this paper we give an efficient approximation scheme for Max-Cut on interval graphs, the intersection graphs of intervals on the real line, and we show, by the same construction, that deleting a suitably chosen small fraction of the edges of an interval graph yields a graph whose clique-width is bounded in terms of that fraction alone. We also show that, unless $\Pclass=\NP$, approximation schemes for Max-Cut do not extend from interval graphs to chordal graphs.

The complexity of Max-Cut on interval graphs was a long-standing open problem; according to Adhikary, Bose, Mukherjee, and Roy, it was probably first mentioned as open in Johnson's NP-completeness column of 1985~\cite{johnson1985}. They settled it in 2021 by proving that the problem is NP-complete~\cite{adhikary2021,adhikary2023}. It remains NP-complete for interval graphs of interval count four, that is, graphs with an interval representation that uses only four distinct lengths~\cite{figueiredo2024count}, and a preprint of Barsukov and Roy proves the same for interval count two~\cite{br2022}. For unit interval graphs, in which all intervals have the same length, the complexity is open~\cite{ahn2025,glp2025}; two polynomial-time algorithms proposed for this case were later reported to be incorrect (see~\cite{adhikary2021,kmn2021}). Adhikary, Bose, Mukherjee, and Roy also proposed the search for approximation algorithms for Max-Cut on interval graphs, and asked whether the ratio $\agw$ can be improved on interval graphs or unit interval graphs~\cite[Section~4]{adhikary2021}. Ahn, DeHaan, Kim, and Lee recently showed that it can be improved on interval graphs: they gave a polynomial-time algorithm with ratio $\agw+10^{-34}$ in expectation~\cite[Theorem~1]{ahn2025}. Approximation schemes exist for special cases: the general schemes for dense graphs apply to dense interval graphs~\cite{akk1999,fdlv1996}, and for unit interval graphs, George, Louis, and Paul state that the method of~\cite{jkls2005}, which was formulated for Max-Bisection, yields an approximation scheme for Max-Cut on unit ball graphs, the intersection graphs of balls of equal radius in $\mathbb R^k$, for fixed $k$ and given a geometric representation; these include the unit interval graphs~\cite{glp2025}.

\subsection{Results}\label{sec:results}

An \emph{interval representation} of a graph $G$ assigns to every vertex a closed interval of the real line such that two vertices are adjacent exactly when their intervals intersect. Our main result is that Max-Cut on interval graphs admits an efficient polynomial-time approximation scheme (EPTAS), that is, an algorithm that, for every $\eps>0$, computes a set $S$ with $\cut_G(S)\ge(1-\eps)\mc(G)$ in time $f(\eps)\cdot n^{O(1)}$ for some function $f$. Our algorithms take an interval representation as part of the input. This is no restriction: the linear-time interval graph test of Booth and Lueker~\cite{bl1976} arranges the maximal cliques of the graph in an order in which the cliques that contain any given vertex are consecutive, and assigning to every vertex the range of positions of these cliques gives an interval representation. We consider unweighted graphs: with nonnegative edge weights, Max-Cut on interval graphs is as hard to approximate as on general graphs, since complete graphs are interval graphs. We measure time in the word-RAM model, assuming that two endpoints of the representation can be compared in constant time (\cref{sec:prelims}). Max-Bisection asks for a set $S$ with $|S|=\lfloor n/2\rfloor$ that maximizes $\cut_G(S)$; we write $\mb(G)$ for the maximum.

\newcommand{\thmEptasText}{There is a deterministic algorithm that, given an interval representation of an unweighted $n$-vertex graph $G$ and a number $\eps\in(0,1)$, computes a set $S\subseteq V(G)$ with $\cut_G(S)\ge(1-\eps)\mc(G)$ in time $2^{O(\eps^{-2}\log^3(2/\eps))}\cdot n^2$. Within the same time bound it can also compute a set $S$ with $|S|=\lfloor n/2\rfloor$ and $\cut_G(S)\ge(1-\eps)\mb(G)$.}
\begin{theorem}\label{thm:eptas}
\thmEptasText
\end{theorem}

We prove \cref{thm:eptas} in \cref{sec:algorithm}. It replaces the fixed ratio $\agw+10^{-34}$ of~\cite{ahn2025} by $1-\eps$ for every $\eps>0$. The price is the dependence of the running time on $\eps$, which is exponential in $\eps^{-2}$ up to logarithmic factors in the exponent, and some superpolynomial dependence on $1/\eps$ is unavoidable unless $\Pclass=\NP$: since $\mc(G)$ is an integer smaller than $n^2$, an algorithm whose running time is polynomial in both $n$ and $1/\eps$ (an FPTAS) would solve the NP-hard problem exactly when run with $\eps=n^{-2}$. Among the standard types of approximation schemes, an EPTAS is therefore the strongest one possible. The same holds for Max-Bisection, which is NP-hard on interval graphs as well: adding $n$ isolated vertices to an $n$-vertex interval graph $G$ gives an interval graph $G'$ with $\mb(G')=\mc(G)$, so the NP-hardness of Max-Cut~\cite{adhikary2021} carries over. We have not tried to optimize the dependence on $n$. For unit interval graphs, George, Louis, and Paul state a faster bound: the method of~\cite{jkls2005}, combined with the dense-graph algorithm of Mathieu and Schudy~\cite{ms2008}, gives an approximation scheme for Max-Cut on unit ball graphs in fixed dimension, given a geometric representation, and hence on unit interval graphs, with running time $O(m)+2^{O(1/\eps^2)}$~\cite[Remark~4]{glp2025}.

The algorithm and a structural theorem about interval graphs rest on one construction. The \emph{clique-width} $\cw(H)$ of a graph $H$ is the least number of labels needed to construct $H$ from single labeled vertices by disjoint unions, by adding all edges between the vertices of two labels, and by relabeling~\cite{co2000}; the formal definition is in \cref{sec:prelims}. Graphs of bounded clique-width include dense graphs such as cliques, and Max-Cut, among many other problems, can be solved on them by dynamic programming over a clique-width expression~\cite{fomin2014}.

\newcommand{\thmStructureText}{For every interval graph $G$ with $m$ edges and every $\eps\in(0,1)$, there is a spanning subgraph $H$ of $G$ that is an interval graph and satisfies $|E(G)\setminus E(H)|\le\eps m$ and $\cw(H)=O(\eps^{-2}\log^2(2/\eps))$. Given an interval representation of $G$, such a graph $H$ and a clique-width expression for $H$ of this width can be computed deterministically in time polynomial in $n$ and $1/\eps$.}
\begin{theorem}\label{thm:structure}
\thmStructureText
\end{theorem}

We prove \cref{thm:structure} in \cref{sec:cliquewidth}; that $H$ is an interval graph is shown in \cref{sec:rounding}. Deleting edges cannot be avoided: unit interval graphs already have unbounded clique-width~\cite{gr2000}. The number of deleted edges is measured against $m$, not against $n^2$, so the theorem is meaningful for sparse graphs. Since every edge contributes at most one to a cut, $\cut_H(S)\le\cut_G(S)\le\cut_H(S)+\eps m$ holds for all sets $S$ simultaneously, and since $\mc(G)\ge m/2$, any $(1-\eps)$-approximate cut of $H$ is a $(1-3\eps)$-approximate cut of $G$.

Exact dynamic programming over a clique-width expression of width $w$ solves Max-Cut in time $n^{O(w)}$. Fomin, Golovach, Lokshtanov, and Saurabh proved that an $f(w)n^{o(w)}$ algorithm on all such inputs would refute the Exponential Time Hypothesis~\cite[Theorem~4.1]{fomin2014}.\footnote{In October 2026, OpenAI~\cite{openai2026eth} and Anthropic~\cite{anthropic2026eth} released AI-generated manuscripts and Lean formalizations giving deterministic $3$-SAT algorithms on $n$ variables in time $2^{o(n)}$ times a polynomial in the input length. Their stated results refute ETH, but leave the conditional implication of~\cite{fomin2014} intact. We cite it only as a historical comparison; none of our results assumes ETH.}
Combined with \cref{thm:structure}, the exact algorithm gives an approximation scheme, but not an efficient one. Lampis gave a randomized algorithm that, given a clique-width expression of width $w$, finds a cut with at least $\mc(H)/(1+\eps)$ edges with high probability in time $(\log n/\eps)^{O(w)}n^{O(1)}$~\cite[full version, Theorem~1]{lampis2014}. Since $(\log n)^w\le w^{2w}+n^2$, combining his algorithm with \cref{thm:structure} already yields a randomized EPTAS for Max-Cut on interval graphs. We prove \cref{thm:eptas} instead by a direct dynamic program over the construction behind \cref{thm:structure}, without a clique-width expression. This construction divides the line into a hierarchy of \emph{cells} by repeated halving and stores every retained rounded interval at the smallest cell that contains it. The hierarchy has more structure than bounded clique-width: every such interval is adjacent either to all or to none of the intervals inside any cell that lies sufficiently many levels below its own (\cref{sec:approaches,sec:ov-delay}). The state of the dynamic program at a cell therefore consists only of rounded counts for the few groups of twin vertices that are stored fewer than that many levels above the cell and meet it, and of the exact number of chosen vertices inside the cell. The count of each group is restricted to a few allowed values and chosen once, at the cell where the group is stored, so rounding errors do not accumulate and no randomized rounding is needed; and since the exact size of the chosen set is part of the state, the algorithm also handles the exact balance constraint of Max-Bisection.

\paragraph{Concurrent work.}
Independently and concurrently, DeHaan and Lee~\cite{dl2026}
obtained a related structural result showing that every interval
graph with $m$ edges can be transformed, by deleting at most
$\eps m$ edges, into a graph of rank-width bounded in terms of
$\eps$ alone. They also develop approximation schemes for
Max-2-CSPs on graphs of bounded rank-width, with applications
to interval graphs. Our structural theorem gives the explicit
bound $O(\eps^{-2}\log^2(2/\eps))$ on clique-width, and hence
also on rank-width~\cite{os2006}, while preserving the
interval-graph property. Moreover, the additional structure of
our construction yields a direct deterministic EPTAS for both
Max-Cut and Max-Bisection.

Our third result shows that, unless $\Pclass=\NP$, Max-Cut admits no approximation scheme on chordal graphs, a class that contains the interval graphs. A graph is \emph{chordal} if it has no induced cycle of length at least four; interval graphs are chordal, and they are exactly the chordal graphs without an asteroidal triple, that is, without three pairwise nonadjacent vertices each two of which are joined by a path that avoids the neighbors of the third~\cite{lb1962}. A \emph{split graph} is a graph whose vertex set can be partitioned into a clique and an independent set; split graphs are chordal as well.

\newcommand{\thmHardnessText}{For every $\eps>0$, it is NP-hard to approximate Max-Cut on unweighted chordal graphs within a factor of $84/85+\eps$. This holds even for graphs obtained from a split graph by adding vertices of degree two whose two neighbors are adjacent.}
\begin{theorem}\label{thm:hardness}
\thmHardnessText
\end{theorem}

We prove \cref{thm:hardness} in \cref{sec:hardness}. 
Bodlaender and Jansen showed that Max-Cut is NP-hard on split graphs, and hence on chordal graphs~\cite{bj2000}; as Ahn, DeHaan, Kim, and Lee remark, this and the hardness results for interval graphs do not imply hardness of approximation~\cite{ahn2025}. For split graphs, Ahn, DeHaan, Kim, and Lee gave a polynomial-time algorithm with ratio $\agw+10^{-16}$ in expectation~\cite[Theorem~2]{ahn2025}, and they showed that, under the Small Set Expansion Hypothesis and with randomized reductions, there is a constant $c>0$ such that no polynomial-time algorithm achieves the ratio $1-c$~\cite[Theorem~3]{ahn2025}; this hypothesis of Raghavendra and Steurer asserts that the edge expansion of small vertex sets is hard to approximate, and it implies the Unique Games Conjecture~\cite{rs2010}. To their knowledge, their result was the first hardness of approximation for Max-Cut on a class of perfect graphs, the graphs in which every induced subgraph has chromatic number equal to its clique number, a class that contains all chordal graphs. \Cref{thm:hardness} assumes only $\Pclass\ne\NP$, at the price of a class slightly larger than the split graphs; to our knowledge, it is the first NP-hardness of approximation for Max-Cut on a class of perfect graphs.

Together with~\cite[Theorem~3]{ahn2025}, \cref{thm:eptas} separates interval graphs from split graphs, and hence from chordal graphs, with respect to approximation schemes under the Small Set Expansion Hypothesis; \cref{thm:eptas,thm:hardness} give this separation from chordal graphs assuming only $\Pclass\ne\NP$. On chordal graphs, the best ratio achievable in polynomial time remains open: the algorithm of Goemans and Williamson achieves $\agw$ in expectation, the tradeoff between triangle packings and the maximum cut behind the algorithms of Ahn, DeHaan, Kim, and Lee fails on some chordal graphs~\cite[Theorem~13]{ahn2025}, and \cref{thm:hardness} rules out, for deterministic algorithms, every ratio above $84/85\approx0.988$ unless $\Pclass=\NP$. \Cref{tab:landscape} summarizes the approximability of Max-Cut on the classes discussed here.

\begin{table}[t]
\centering
\renewcommand{\arraystretch}{1.15}
\begin{tabular}{@{}>{\raggedright\arraybackslash}p{2.0cm}>{\raggedright\arraybackslash}p{4.0cm}>{\raggedright\arraybackslash}p{2.3cm}>{\raggedright\arraybackslash}p{6.8cm}@{}}
\toprule
Class & Algorithms before this paper & This paper & Hardness\\
\midrule
all graphs & ${\agw}$~\cite{goemans1995} & & ${16/17+\eps}$~\cite{hastad2001,tssw2000,cst2001}; ${\agw+\eps}$ under the Unique Games Conjecture (classical)~\cite{kkmo2007,moo2010}\\
chordal & ${\agw}$~\cite{goemans1995} & & ${84/85+\eps}$ (\cref{thm:hardness}), even for split graphs with added vertices of degree two whose two neighbors are adjacent\\
split & ${\agw+10^{-16}}$~\cite{ahn2025} & & ${1-c}$ for some $c>0$, under the Small Set Expansion Hypothesis with randomized reductions~\cite{ahn2025}; exact solution NP-hard~\cite{bj2000}\\
interval & ${\agw+10^{-34}}$~\cite{ahn2025} & deterministic EPTAS (\cref{thm:eptas}) & exact solution NP-hard~\cite{adhikary2021}\\
unit interval & approximation scheme, time ${O(m)+2^{O(1/\eps^2)}}$, as stated in~\cite{glp2025} & & open whether exact solution is NP-hard\\
\bottomrule
\end{tabular}
\caption{Approximation ratios for Max-Cut achievable in polynomial time, and ratios within which approximation is NP-hard for every $\eps>0$, unless $\Pclass=\NP$ or unless stated otherwise. Ratios of randomized algorithms, such as those of~\cite{goemans1995} and~\cite{ahn2025}, hold in expectation. Chordal graphs contain split graphs and interval graphs, which are incomparable, and interval graphs contain unit interval graphs. Footnote~\ref{fn:ugc-update} records the October 2026 UGC update.}\label{tab:landscape}
\end{table}

\subsection{Why known approaches do not suffice}\label{sec:approaches}

Approximation schemes for Max-Cut on special graph classes follow, broadly, one of three routes, and each stops short of interval graphs.

\emph{Density.} For graphs with $\Omega(n^2)$ edges, sampling~\cite{akk1999,fdlv1996} and the weak regularity lemma of Frieze and Kannan~\cite{frieze1999} compute cuts whose size is within an additive $\eps n^2$ of the optimum, which is a $(1-O(\eps))$-approximation when $m=\Omega(n^2)$. The regularity lemma of Fox, Pach, and Suk for semi-algebraic graphs~\cite{fox2016}, a class that contains interval graphs, partitions the vertices into $(1/\eps)^{O(1)}$ parts whose sizes differ by at most one, such that all but an $\eps$-fraction of the pairs of parts are complete or empty; since the parts have almost equal sizes, it, too, controls the error only up to about $\eps n^2$ pairs of vertices. Interval graphs can be sparse, however: a family of pairwise disjoint short intervals together with a few long ones can have $m=O(n)$, and then an additive error of $\eps n^2$ exceeds the optimum. \Cref{thm:structure} bounds the error by $\eps m$ instead, for every interval graph.

\emph{Bounded width.} The algorithms for graphs given with a clique-width expression~\cite{fomin2014,lampis2014} do not apply directly, because interval graphs have unbounded clique-width~\cite{gr2000}. For planar graphs, the layering technique of Baker~\cite{baker1994}, in the edge-deletion form used for Max-Bisection by Jansen, Karpinski, Lingas, and Seidel~\cite{jkls2005}, deletes a small fraction of the edges to obtain pieces of bounded treewidth. For interval graphs, no deletion of $\eps m$ edges bounds the treewidth: interval graphs contain cliques of every size, and a graph of treewidth $t$ on $k$ vertices has fewer than $tk$ edges. \Cref{thm:structure} therefore approximates the input graph by a graph of bounded clique-width, a measure under which cliques are simple, and the dynamic program of \cref{thm:eptas} uses the additional structure of the graph that we construct.

\emph{Geometric decomposition.} The shifting technique of Hochbaum and Maass~\cite{hm1985}, in its form for packing problems, partitions an instance along a grid of lines, chosen among a few shifted copies, discards the objects that the lines cut, and solves the pieces separately; averaging over the shifts shows that few objects are lost. This single-scale form uses a common size scale for the objects, which sets the grid spacing, and interval graphs with arbitrary lengths have no common length scale. Under a uniformly random shift, grid lines at spacing $D$ meet an interval of length $L$ with probability $\min\{1,L/D\}$, so a spacing that rarely cuts long intervals leaves cells that contain arbitrarily many short intervals in arbitrary positions, and a spacing that keeps cells simple cuts many long intervals, each of which may carry many edges. For disk graphs with arbitrary diameters, Erlebach, Jansen, and Seidel overcame the lack of a common scale by classifying the disks into levels by size and applying the shifting strategy on all levels of a recursive subdivision simultaneously, so that every disk is tested only against the lines of its own level~\cite{ejs2005}; Chan gave a variant based on shifted quadtrees~\cite{chan2003}. These schemes solve problems such as maximum independent set, for which the dynamic program needs to remember only a bounded number of objects of a solution at each cell. For Max-Cut every vertex counts, and a cell can meet arbitrarily many intervals of every scale, so the interaction of long intervals with the many short intervals below them must be summarized instead.

Our construction combines a shifted hierarchy of cells obtained by repeated halving, the one-dimensional analogue of the shifted quadtrees in Arora's scheme for Euclidean problems~\cite{arora1998}, with three further ingredients. First, we measure lengths in \emph{rank coordinates}, obtained by replacing every endpoint by its rank; this normalization is standard, and what we use is that in these coordinates the length of an interval is at most twice its degree plus one, so that rounding an interval to a grid adapted to its own length can be paid for by its own edges. Second, as in the multi-level shifting of~\cite{ejs2005,chan2003}, every interval is tested against grid lines only at its own scale; the intervals that are cut lose their edges, and we choose the shift that minimizes their total degree. Third, after storing every rounded interval at the smallest cell that contains it, an interval acts uniformly on every cell that lies at least $\tau=O(\log(2/\eps))$ levels below its own: it is adjacent to all intervals stored in the subtree of such a cell, or to none. Long intervals are therefore not removed but counted. \Cref{sec:overview} explains how these ingredients yield \cref{thm:eptas,thm:structure}.

\subsection{Further related work}\label{sec:related}

George, Louis, and Paul show that the standard semidefinite relaxation yields, in nearly linear time, a ratio $\agw+\Omega(1)$ in expectation on graphs with $\Omega(m)$ edge-disjoint triangles, a criterion met by unit ball graphs in fixed dimension of sufficiently large average degree, and in particular by such unit interval graphs~\cite{glp2025}. The algorithms of Ahn, DeHaan, Kim, and Lee combine a tradeoff between edge-disjoint triangle packings and the maximum cut with an analysis of a perturbed Goemans--Williamson algorithm~\cite{ahn2025}. Under a boundedness condition, Coja-Oghlan, Cooper, and Frieze give sparse regularity approximations whose error in the cut norm, that is, in the numbers of edges between pairs of vertex sets, is proportional to the number of edges~\cite{coja2010}; their approximation is a sum of cut matrices, whereas \cref{thm:structure} applies to every interval graph and produces an edge-deletion subgraph with an explicit width expression. In this respect \cref{thm:structure} resembles hyperfiniteness of graphs of bounded degree, where removing $\eps n$ suitably chosen edges leaves components of bounded size~\cite{ns2013}, and fragility with respect to bounded treewidth, which implies that removing a small fraction of suitably chosen vertices leaves a graph of bounded treewidth~\cite{dvorak2016}; in \cref{thm:structure}, the number of removed edges is measured against $m$, and the target is bounded clique-width, which allows large cliques.

\subsection{Organization}

\Cref{sec:overview} sketches the proofs of all three theorems. \Cref{sec:prelims} fixes notation. \Cref{sec:structure} constructs the edited graph $H$ and its hierarchical description, which both \cref{thm:eptas} (\cref{sec:algorithm}) and \cref{thm:structure} (\cref{sec:cliquewidth}) use; these two sections can be read independently of each other. \Cref{sec:hardness} proves \cref{thm:hardness} and does not depend on the earlier technical sections.

\section{Technical overview}\label{sec:overview}

This section sketches the proofs of \cref{thm:eptas,thm:structure,thm:hardness}; the complete arguments are in \cref{sec:structure,sec:algorithm,sec:cliquewidth,sec:hardness}. Throughout, $\sigma=2^{-b}$ is a power of two with $\sigma\le1/16$, which we eventually choose as $\Theta(\eps)$, and logarithms are binary.

Vertices with the same interval are true twins, that is, adjacent vertices with the same neighbors outside the pair; a cut sees a set of true twins only through the number of its vertices on each side. A direct sweep-line dynamic program would keep one count for each distinct interval crossing the current point, and there may be $\Theta(n)$ such intervals. Since $\mc(G)\ge m/2$, we can afford to delete $O(\sigma m)$ edges to reduce their number. We present the three steps of the construction in the order in which they become necessary: rounding (\cref{sec:ov-rounding}) leaves too many scales; a hierarchy of cells removes them, since long intervals act uniformly on small cells, provided that no short interval is stored at a long cell (\cref{sec:ov-delay}); and a shift of the hierarchy, at the cost of the edges of a few intervals, ensures this (\cref{sec:ov-shift}).

\subsection{Rounding each interval at its own scale}\label{sec:ov-rounding}

Rounding all endpoints to one common grid does not suffice: a grid that is coarse enough to make long intervals simple destroys the short ones, and every destroyed short interval may carry all its edges. We therefore round every interval to a grid adapted to its own length, and we measure length so that the loss is paid for by the edges of the interval itself.

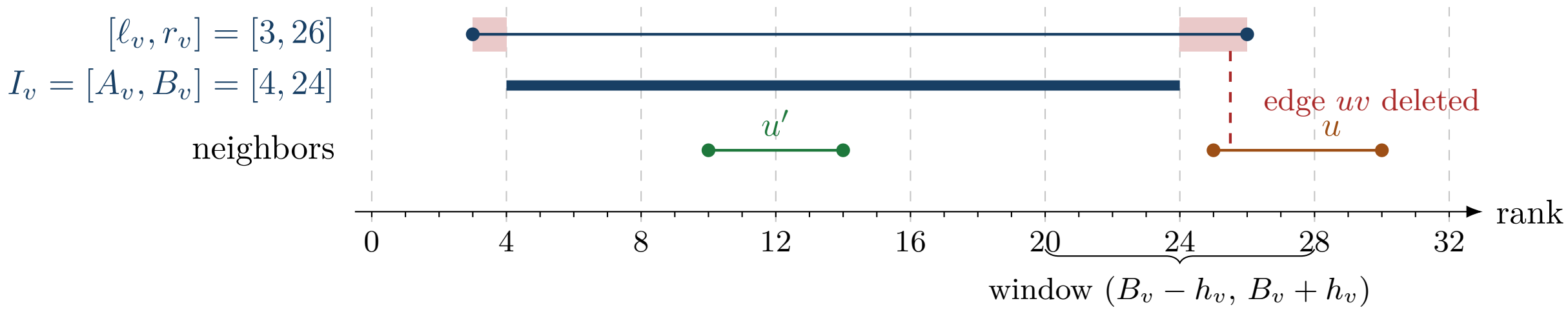
\begin{figure}[htbp]
\centering
\begin{tikzpicture}[x=0.345cm,y=0.7cm,>=Latex,font=\small]
\foreach \x in {0,4,...,32} {\draw[gray!45,dashed] (\x,-0.15)--(\x,3.0);}
\draw[->] (-0.5,0)--(33,0) node[right]{rank};
\foreach \x in {0,...,32} {\draw (\x,0)--(\x,-0.08);}
\foreach \x in {0,4,...,32} {\node[below,font=\footnotesize] at (\x,-0.1) {\x};}
\fill[figred!25] (3,2.35) rectangle (4,2.85);
\fill[figred!25] (24,2.35) rectangle (26,2.85);
\draw[figblue,thick] (3,2.6)--(26,2.6);
\fill[figblue] (3,2.6) circle (2pt) (26,2.6) circle (2pt);
\node[anchor=east,figblue] at (-0.7,2.6) {$[\ell_v,r_v]=[3,26]$};
\draw[figblue,line width=3pt] (4,1.85)--(24,1.85);
\node[anchor=east,figblue] at (-0.7,1.85) {$I_v=[A_v,B_v]=[4,24]$};
\node[anchor=east] at (-0.7,0.9) {neighbors};
\draw[figgreen,thick] (10,0.9)--(14,0.9);
\fill[figgreen] (10,0.9) circle (2pt) (14,0.9) circle (2pt);
\node[figgreen,above] at (12,0.92) {$u'$};
\draw[figwarm,thick] (25,0.9)--(30,0.9);
\fill[figwarm] (25,0.9) circle (2pt) (30,0.9) circle (2pt);
\node[figwarm,above] at (28.5,0.92) {$u$};
\draw[figred,dashed,thick] (25.5,1.0)--(25.5,2.5);
\node[figred,anchor=west,font=\footnotesize] at (26.1,1.6) {edge $uv$ deleted};
\draw[decorate,decoration={brace,amplitude=4pt,mirror}] (20,-0.55)--(28,-0.55) node[midway,below=4pt,font=\footnotesize]{window $(B_v-h_v,\,B_v+h_v)$};
\end{tikzpicture}
\caption{Rounding in rank coordinates, drawn with $\sigma=1/4$ for legibility; the construction uses $\sigma\le1/16$. The interval of $v$ has rank length $L_v=23$, scale $s_v=16$ and grid spacing $h_v=\sigma s_v=4$ (dashed lines); rounding removes the shaded ends. The neighbor $u$ has $h_u=1$ and does not move, so the edge $uv$ is deleted, while $u'$ remains adjacent to $v$. A deleted edge $uv$ with $h_u\le h_v$ always has an endpoint of $u$ within distance less than $h_v$ of an endpoint of $I_v$; here the endpoint $25$ of $u$ lies in the window around $B_v$ marked by the brace.}\label{fig:charging}
\end{figure}

Replace the $2n$ endpoints of the representation by their ranks $1,\dots,2n$, placing left endpoints before right endpoints at equal positions; this does not change the graph. Let $\ell_v<r_v$ be the ranks of the endpoints of $v$ and let $L_v=r_v-\ell_v$ be its \emph{rank length}. Each of the $L_v-1$ ranks strictly between $\ell_v$ and $r_v$ is an endpoint of an interval that meets the interval of $v$, and each neighbor of $v$ has two endpoints. Hence
\begin{equation}\label{eq:ov-rank}
L_v\le2\deg_G(v)+1 .
\end{equation}
In rank coordinates, length is therefore paid for by degree, whatever the density of the graph. Let $s_v$ be the largest power of two with $s_v\le L_v$, the \emph{scale} of $v$, and let $h_v=\max\{1,\sigma s_v\}$, its \emph{grid spacing}. We move $\ell_v$ up and $r_v$ down to the nearest multiples of $h_v$, obtaining a \emph{rounded interval} $I_v=[A_v,B_v]\subseteq[\ell_v,r_v]$, whose length lies between $s_v/2$ and $2s_v$. (The rounding is done after the shift of the coordinates chosen in \cref{sec:ov-shift}.) Since intervals only shrink, edges can only be deleted. If an edge $uv$ is deleted and $h_u\le h_v$, then, because both endpoints moved by less than $h_v$, an endpoint of $u$ lies within distance less than $h_v$ of an endpoint of $I_v$, as in \cref{fig:charging}; we charge the edge to $v$. Ranks are distinct integers, and distinct edges charged to $v$ have distinct such endpoints of other intervals, so fewer than $4h_v$ deleted edges are charged to $v$. Intervals with $h_v=1$ do not move, and for the others $h_v=\sigma s_v\le\sigma L_v\le3\sigma\deg_G(v)$ by~\eqref{eq:ov-rank}. Summing $4h_v\le12\sigma\deg_G(v)$ over the vertices with $h_v>1$, to which all deleted edges are charged, bounds the rounding loss by $24\sigma m$. Stretching the intervals instead would cost about as much; we shrink them because then every edit of the construction is a deletion, so that the edited graph $H$ is a spanning subgraph of $G$.

Rounding alone gives an approximation scheme, but not an efficient one. At most $(2/\sigma+1)^2$ distinct rounded intervals of each scale $s$ contain a given point, since their endpoints are grid points within distance $2s$ of it; but there are up to $\log(2n)$ scales, so a dynamic program that sweeps the line and records a count for each rounded interval through the current point, with counts restricted as in \cref{sec:ov-dp}, has $n^{O(\eps^{-2}\log(2/\eps))}$ states.

\subsection{Long intervals act uniformly on small cells}\label{sec:ov-delay}

Long intervals become simple when they are seen from a region that is short compared with their grid spacing. Let $J$ be a rounded interval with grid spacing $h$, and let $C$ be an open interval between two consecutive multiples of a power of two $q\le h$. The endpoints of $J$ are multiples of $h$, hence of $q$, so neither of them lies in $C$, and $J$ either contains $C$ or misses it. Consider $W$ vertices whose rounded intervals lie inside $C$, of which a cut $S$ contains $t$, and $U$ vertices with such long rounded intervals that meet $C$, of which $S$ contains $R$. Every one of the former is adjacent to every one of the latter, so the number of cut edges between them is
\begin{equation}\label{eq:ov-affine}
t(U-R)+(W-t)R .
\end{equation}
This number depends on the long intervals only through $U$ and $R$, and on the inside of $C$ only through $W$ and $t$, however many long intervals of however many scales there are. Long intervals can therefore be counted rather than described individually, but only from aligned regions, which a sweep line does not provide: the intervals to the right of a point see a long interval through that point only up to its right endpoint.

We use the aligned hierarchy of \cref{sec:hierarchy}, with root $(-2M,2M)$, where $M$ is the least power of two at least $2n+1$. Every cell of length at least two has its two open halves as children. Each nonroot cell lies between consecutive multiples of its length. We store every rounded interval at the smallest cell that contains it, which we call its \emph{anchor}. The rounded interval contains the midpoint of its anchor, since otherwise it would fit into a child. The intervals anchored in the subtree of a cell are exactly the intervals inside the cell, and intervals anchored at incomparable cells lie in disjoint cells, so every edge of $H$ joins two intervals whose anchors lie on a common root-to-leaf path. Vertices with the same scale and the same rounded interval, and hence the same anchor, form a \emph{group}, and we write $I_g$, $s_g$, and $h_g$ for the common rounded interval, scale, and grid spacing of the vertices of a group $g$. A group is a clique of true twins.

\begin{figure}[t]
\centering
\begin{tikzpicture}[x=0.55cm,y=0.56cm,>=Latex,font=\small,
  lab/.style={anchor=west,font=\footnotesize,text width=6.3cm,align=left}]
\fill[figwarm!25] (0,0) rectangle (16,-0.6);
\fill[figwarm!25] (0,-1) rectangle (16,-1.6);
\fill[figwarm!25] (4,-2) rectangle (12,-2.6);
\fill[figwarm!25] (4,-3) rectangle (6,-3.6);
\fill[figwarm!25] (10,-3) rectangle (12,-3.6);
\fill[figgreen!30] (6,-3) rectangle (10,-3.6);
\fill[figgreen!30] (5,-4) rectangle (11,-4.6);
\foreach \lev/\n in {0/1,1/2,2/4,3/8,4/16} {
  \pgfmathsetmacro{\w}{16/\n}
  \foreach \i in {1,...,\n} {
    \pgfmathsetmacro{\xa}{(\i-1)*\w}
    \draw[gray!70] (\xa,-\lev) rectangle ++(\w,-0.6);
  }
  \node[font=\footnotesize] at (-0.7,-\lev-0.3) {\lev};
}
\node[font=\footnotesize,align=center] at (-0.7,0.45) {level};
\draw[figblue,very thick] (0,0) rectangle (16,-0.6);
\draw[figblue,densely dotted] (5,-0.6)--(5,-5.4);
\draw[figblue,densely dotted] (11,-0.6)--(11,-5.4);
\draw[figwarm,dashed] (8,0.2)--(8,-4.8);
\node[figwarm,font=\footnotesize] at (8,0.45) {midpoint of the anchor};
\node[lab,figblue] at (16.3,-0.3) {anchor of $v$, of length $s_v/\sigma$};
\node[lab] at (16.3,-1.3) {level 1: both cells contain an endpoint of $I_v$};
\node[lab] at (16.3,-3.3) {level $\tau-1=3$: two cells contain an endpoint};
\node[lab] at (16.3,-4.3) {level $\tau=4$: every cell lies in $I_v$ or outside it};
\draw[line width=2pt,figblue] (5,-5.4)--(11,-5.4);
\fill[figblue] (5,-5.4) circle (2.2pt) (11,-5.4) circle (2.2pt);
\node[figblue] at (8,-5.8) {$I_v=[5,11]$};
\draw[gray] (0,-6.3)--(16,-6.3);
\foreach \x in {0,...,16} {\draw (\x,-6.3)--(\x,-6.42) node[below,font=\scriptsize]{\x};}
\end{tikzpicture}
\caption{An anchor and the cells up to $\tau$ levels below it, drawn with $\sigma=1/4$, and hence $\tau=2\log(1/\sigma)=4$, for legibility; the construction uses $\sigma\le1/16$. The vertex $v$ has scale $s_v=4$ and grid spacing $h_v=\max\{1,\sigma s_v\}=1$. Its rounded interval $I_v=[5,11]$ contains the midpoint $8$ of the cell $(0,16)$, which is therefore the anchor of $v$; this anchor has the largest length, $s_v/\sigma=16$, that the construction permits (\cref{sec:ov-shift}). Orange cells contain an endpoint of $I_v$, green cells lie in $I_v$, and white cells are disjoint from $I_v$. From level $\tau$ on, the cells have length at most $h_v$, so $I_v$ contains every cell that it meets, and $v$ is then adjacent to every vertex anchored in the subtree of that cell.}\label{fig:anchor}
\end{figure}
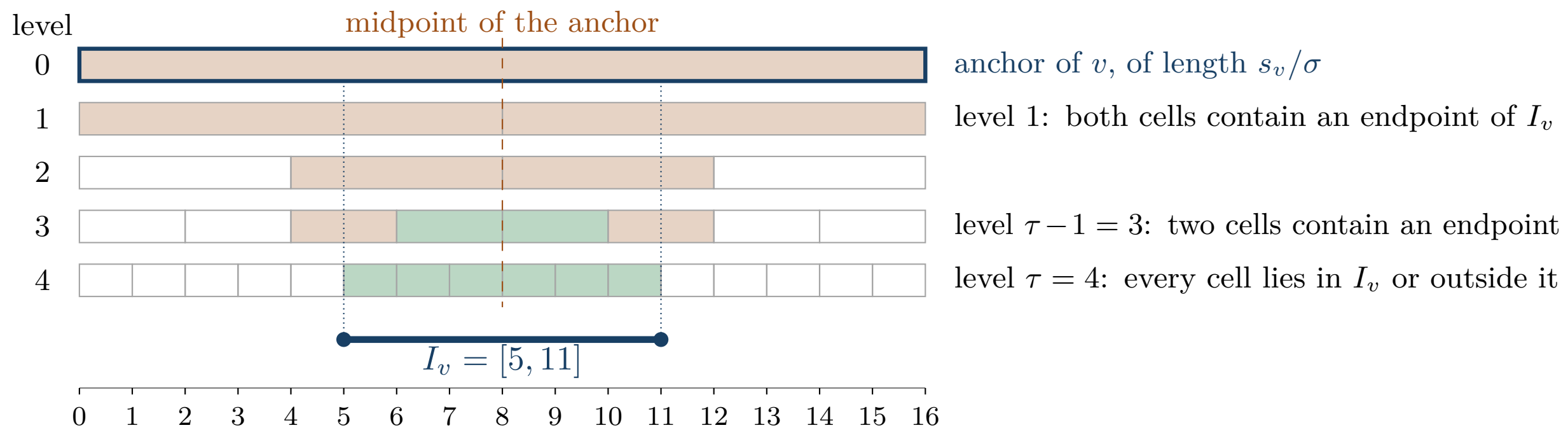

Suppose now that the anchor of every interval $v$ has length at most $s_v/\sigma$. Then a group $g$ anchored at a cell of length $D$ has $s_g\ge\sigma D$ and hence $h_g\ge\max\{1,\sigma^2D\}$, one factor $\sigma$ coming from the anchor and one from the grid. Every cell $C$ that lies $\tau=2b$ or more levels below the anchor has length at most $\sigma^2D\le h_g$, so $I_g$ either contains $C$ or misses it (\cref{lem:old-universal}); in the first case, every vertex anchored in the subtree of $C$ is adjacent to every vertex of $g$ (\cref{fig:anchor}). Among the groups that are anchored at a proper ancestor of $C$ and meet $C$, we call those anchored fewer than $\tau$ levels above $C$ \emph{recent} for $C$ and the others \emph{old} for $C$; old groups contain $C$, and their edges to the subtree of $C$ are given by~\eqref{eq:ov-affine}. Moreover, a group anchored at a cell of length $D$ has one of the $b+1$ scales between $\sigma D$ and $D$, and for each scale $s$, both endpoints of its rounded interval are multiples of $\max\{1,\sigma s\}$ within distance $2s$ of the midpoint of the cell. A cell is therefore the anchor of only $O(\sigma^{-2}\log(1/\sigma))$ groups (\cref{lem:groups-per-anchor}), and it has only $O(\sigma^{-2}\log^2(1/\sigma))$ recent groups, which are anchored at the $\tau-1$ cells just above it. Neither bound depends on $n$.

The assumption does not hold by itself: an interval that contains the midpoint of a long cell is anchored there, however short it is. For instance, a short interval through the root midpoint can remain nonuniform for $\Theta(\log n)$ levels, and the root may anchor intervals of all scales.

\subsection{Shifting the hierarchy and rejecting few intervals}\label{sec:ov-shift}

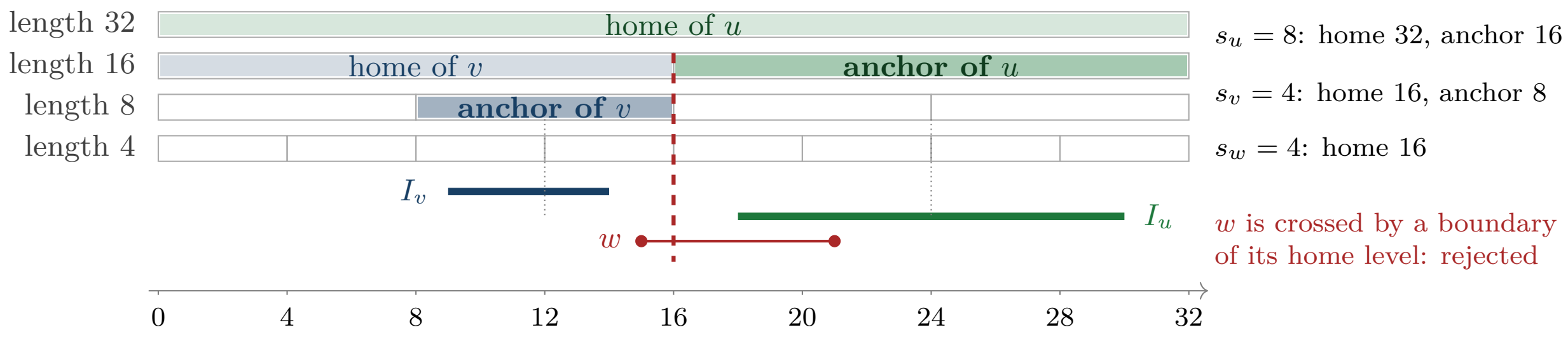
\begin{figure}[t]
\centering
\begin{tikzpicture}[x=0.335cm,y=0.43cm,font=\footnotesize]
\foreach \lev/\len in {0/32,1/16,2/8,3/4} {
  \pgfmathtruncatemacro{\n}{32/\len}
  \foreach \i in {1,...,\n} {\draw[gray!65] ({(\i-1)*\len},-\lev) rectangle ++(\len,-0.62);}
  \node[anchor=east,gray!50!black] at (-0.3,-\lev-0.31) {length \len};
}
\fill[figgreen!18] (0.06,-0.06) rectangle (31.94,-0.56);
\node[figgreen!60!black] at (16,-0.31) {home of $u$};
\fill[figblue!18] (0.06,-1.06) rectangle (15.94,-1.56);
\node[figblue] at (8,-1.31) {home of $v$};
\fill[figgreen!40] (16.06,-1.06) rectangle (31.94,-1.56);
\node[figgreen!50!black] at (24,-1.31) {\textbf{anchor of $u$}};
\fill[figblue!40] (8.06,-2.06) rectangle (15.94,-2.56);
\node[figblue] at (12,-2.31) {\textbf{anchor of $v$}};
\draw[figred,very thick,dashed] (16,-1.0)--(16,-6.05);
\draw[->,gray] (-0.3,-6.75)--(32.6,-6.75);
\foreach \x in {0,4,...,32} {\draw[gray] (\x,-6.75)--(\x,-6.85) node[below,font=\scriptsize,text=black]{\x};}
\draw[figblue,line width=2.2pt] (9,-4.35)--(14,-4.35); \node[figblue,anchor=east] at (8.8,-4.35) {$I_v$};
\draw[figgreen,line width=2.2pt] (18,-4.95)--(30,-4.95); \node[figgreen,anchor=west] at (30.2,-4.95) {$I_u$};
\draw[figred,thick] (15,-5.55)--(21,-5.55); \fill[figred] (15,-5.55) circle (1.8pt) (21,-5.55) circle (1.8pt); \node[figred,anchor=east] at (14.8,-5.55) {$w$};
\foreach \x in {12,24} {\draw[gray,densely dotted] (\x,-2.6)--(\x,-4.95);}
\node[anchor=west,align=left,font=\scriptsize] at (32.4,-0.6) {$s_u=8$: home $32$, anchor $16$};
\node[anchor=west,align=left,font=\scriptsize] at (32.4,-2.0) {$s_v=4$: home $16$, anchor $8$};
\node[anchor=west,align=left,font=\scriptsize] at (32.4,-3.3) {$s_w=4$: home $16$};
\node[figred,anchor=west,align=left,font=\scriptsize] at (32.4,-5.55) {$w$ is crossed by a boundary\\of its home level: rejected};
\end{tikzpicture}
\caption{Homes and anchors, drawn with $\sigma=1/4$ for legibility; the construction uses $\sigma\le1/16$. The notes on the right give the lengths of the home and the anchor. Every interval is tested only against the cell boundaries of the level of length $s_v/\sigma$, and dotted lines mark the midpoints of the two anchors. The vertex $v$ has $[\ell_v,r_v]=I_v=[9,14]$; its home is $(0,16)$, and its anchor is $(8,16)$, whose midpoint $12$ lies in $I_v$. The vertex $u$ has $[\ell_u,r_u]=[17,30]$ and $h_u=2$, so $I_u=[18,30]$, with home $(0,32)$ and anchor $(16,32)$; $I_u$ contains the boundaries $20$, $24$, and $28$ of finer levels, which is allowed, since $u$ is tested only at its home level. The vertex $w$, with $[\ell_w,r_w]=[15,21]$, is crossed by the boundary $16$ of its home level and is rejected.}\label{fig:homes}
\end{figure}
We obtain the assumption of \cref{sec:ov-delay} by testing every interval against the cell boundaries at the level of its own size. Before rounding, an interval $v$ has the root as its \emph{home} if $s_v/\sigma\ge4M$. Otherwise its home is the cell of length $s_v/\sigma$ containing its closed interval; if a boundary at that level belongs to the interval, $v$ is \emph{rejected} and becomes isolated in $H$. The rounded interval of every other vertex lies inside its home, so its anchor is a descendant of the home and has length at most $s_v/\sigma$ (\cref{fig:homes}). For a nonroot home, boundaries of longer cells are multiples of $s_v/\sigma$ and cannot meet the interval either; boundaries of shorter cells may, which is why the anchor can be a proper descendant of the home.

The hierarchy is shifted by an integer $\theta$, that is, all coordinates are shifted by $-\theta$ before rounding, so that rounding grids and cell boundaries are multiples of powers of two in the same coordinates. For a uniformly random shift, $v$ is rejected with probability at most $(L_v+1)/(s_v/\sigma)\le2\sigma$, since its interval contains $L_v+1$ integers. A rejected vertex costs its degree, so the expected number of deleted edges is at most $2\sigma\sum_v\deg_G(v)=4\sigma m$, and trying all $O(n)$ shifts finds a shift that achieves this bound deterministically (\cref{lem:shift}). Testing every object only at the level of its own size is the multi-level shifting technique of~\cite{ejs2005,chan2003}, which extends the shifting technique of Hochbaum and Maass~\cite{hm1985} to objects of different sizes; we use a hierarchy obtained by repeated halving, as in~\cite{arora1998}, and we weight rejections by degree. The factor $1/\sigma$ balances two costs: rejections cost $O(\sigma m)$ edges, while an anchor carries only $b+1$ scales and $\tau=2b$ levels suffice.

Homes are long so that rejections are rare, and anchors are short so that intervals are described compactly: a home of length $s/\sigma$ may contain about $\sigma^{-3}$ distinct rounded intervals of scale $s$, whereas the midpoint of the anchor leaves at most $(2/\sigma+1)^2$ for each scale.

\subsection{The dynamic program}\label{sec:ov-dp}

The algorithm of \cref{thm:eptas} runs a dynamic program over the hierarchy, from the leaves to the root. For a cell $C$, let $V_C$ be the set of vertices anchored in the subtree of $C$, and $W(C)=|V_C|$. Every neighbor of a vertex of $V_C$ outside $V_C$ belongs to a recent or an old group of $C$ (\cref{lem:outside-neighbors}). By~\eqref{eq:ov-affine}, the edges between $V_C$ and the old groups depend on $V_C$ only through the number $t$ of its vertices in the cut set $S$, whereas a recent group may end inside $C$, so that the subproblem at $C$ must know how many vertices of the group lie in $S$. Accordingly, the table $F_C(x,t)$ records the largest number of cut edges inside $V_C$ and between $V_C$ and the recent groups of $C$, where the \emph{boundary vector} $x$ prescribes the numbers of vertices of $S$ in the recent groups and $t$ is the exact number of vertices of $S$ in $V_C$; edges to old groups are excluded from $F_C$.

To compute $F_C$, we choose counts for the groups anchored at $C$, the \emph{own} groups of $C$, which are the groups inside $C$ that lie inside neither child, and totals $t_0,t_1$ for the two children. Every recent group of a child is an own or a recent group of $C$, so the boundary vector of each child is read off from $x$ and the chosen counts, and the edges between the vertices anchored in the subtree of a child and the groups that become old exactly at that child are added through~\eqref{eq:ov-affine} (\cref{fig:cell}). No edge joins the subtrees of the two children. In this way every edge of $H$ is counted exactly once: at the anchor of its lower endpoint if the group of the other endpoint is anchored there or recent there, and otherwise through~\eqref{eq:ov-affine}, at the cell where that group becomes old.

We restrict the number of vertices of $S$ in a group $g$ to $O(1/\gamma)$ \emph{allowed counts} for a parameter $\gamma=\Theta(\eps)$, namely $|g|$ and the multiples of $\max\{1,\lfloor\gamma|g|\rfloor\}$ up to $|g|$, so that every integer count lies within $\gamma|g|$ of an allowed one. The count of a group is chosen once, at its anchor, and used unchanged at every cell below. The only error is therefore that an optimal cut need not have allowed counts: rounding its counts moves at most $\gamma|g|$ vertices of each group $g$, all of degree $\deg_H(g)$ in $H$, so the loss is at most $\sum_g\gamma|g|\deg_H(g)=2\gamma|E(H)|$, however deep the hierarchy is. The totals $t$ are kept exact. For every cell, there are $2^{O(\sigma^{-2}\log^2(1/\sigma)\log(1/\gamma))}$ choices of counts, and combining two children with $W_0$ and $W_1$ anchored vertices costs $O((W_0+1)(W_1+1))$ for each choice. A pair of vertices contributes to $W_0W_1$ only at the cell where the paths to their anchors separate, so these products sum to at most $\binom n2$, and the total running time is $2^{O(\eps^{-2}\log^3(2/\eps))}\cdot n^2$.

For Max-Cut, the algorithm returns an allowed set attaining the largest root-table entry. For Max-Bisection, it maximizes over finite entries whose totals $t$ satisfy $|t+z-\lfloor n/2\rfloor|\le\gamma n$ for some $z\in\{0,\ldots,|Z|\}$, where $Z$ is the set of rejected vertices. Rounding a maximum bisection supplies such an entry. The algorithm adds a witnessing number $z$ of rejected vertices and moves at most $\gamma n$ vertices of smallest degree in $G$ from the side exceeding its required size. That side has at least $n/2$ vertices and average degree at most $4m/n$, so balancing loses at most $4\gamma m$. The total loss is at most $28\sigma m+6\gamma m<\eps m/2$ when $28\sigma\le\eps/16$ and $\gamma=\eps/16$. Both optima are at least $m/2$: an unrestricted random cut cuts each edge with probability $1/2$, and a uniformly random bisection with probability at least $1/2$.

\subsection{Few external neighborhoods}\label{sec:ov-width}

\begin{figure}[t]
\centering
\begin{tikzpicture}[x=0.5cm,y=0.41cm,>=Latex,font=\small,
  grp/.style={line width=2.2pt},
  lab/.style={anchor=west,font=\footnotesize}]
\draw[grp,figblue,<->] (-3,5.0)--(19,5.0);
\node[lab,figblue] at (19.3,5.0) {old for $C$};
\draw[grp,figblue!45,<->] (-3,4.2)--(19,4.2);
\node[lab,figblue!70] at (19.3,4.2) {recent, contains $C$};
\draw[grp,figwarm,<-] (-3,3.4)--(5,3.4);
\draw[grp,figwarm,->] (11,3.4)--(19,3.4);
\node[lab,figwarm] at (19.3,3.4) {recent, one endpoint in $C$};
\draw[grp,figviolet,<-] (-3,2.6)--(8,2.6);
\node[lab,figviolet] at (19.3,2.6) {recent for $C$, old for $C_0$};
\draw[thick] (0,1.2) rectangle (16,1.9); \node at (14.9,1.55) {$C$};
\draw (0,0.2) rectangle (8,0.9); \node at (6.9,0.55) {$C_0$};
\draw (8,0.2) rectangle (16,0.9); \node at (14.9,0.55) {$C_1$};
\foreach \x in {5,8,11} {\draw[gray,dashed] (\x,3.4)--(\x,-2.6);}
\node[lab,gray!80!black] at (19.3,1.55) {ancestral endpoint positions};
\draw[gray,dashed] (17.7,1.55)--(19.1,1.55);
\draw[figgreen,thick] (1.5,-0.6)--(4,-0.6);
\draw[figgreen,thick] (5.5,-1.2)--(7.5,-1.2);
\draw[figgreen,line width=1.6pt] (6.5,-1.9)--(10,-1.9);
\draw[figgreen,thick] (9,-0.6)--(12.5,-0.6);
\draw[figgreen,thick] (13.5,-1.2)--(15,-1.2);
\node[lab,figgreen] at (19.3,-0.6) {intervals stored below $C$};
\node[lab,figgreen] at (19.3,-1.9) {own group of $C$};
\end{tikzpicture}
\caption{Schematic view of a cell $C$ with children $C_0$ and $C_1$, of the groups anchored above $C$ that meet $C$ (top), and of the rounded intervals stored in the subtree of $C$ (bottom); arrows indicate that an interval continues beyond the drawing. An old group contains $C$ and is adjacent to every vertex stored below $C$, and its edges to them enter the dynamic program through~\eqref{eq:ov-affine}. A recent group either contains $C$ or has exactly one endpoint inside $C$. These endpoints are the ancestral endpoint positions of $C$ (dashed); the positions of the two endpoints of a stored interval relative to them determine its external neighborhood (\cref{sec:ov-width}). The violet group is recent for $C$ and old for $C_0$, which it contains; its edges to the vertices stored below $C_0$ are added through~\eqref{eq:ov-affine} when the dynamic program passes from $C$ to $C_0$ (\cref{sec:ov-dp}).}\label{fig:cell}
\end{figure}
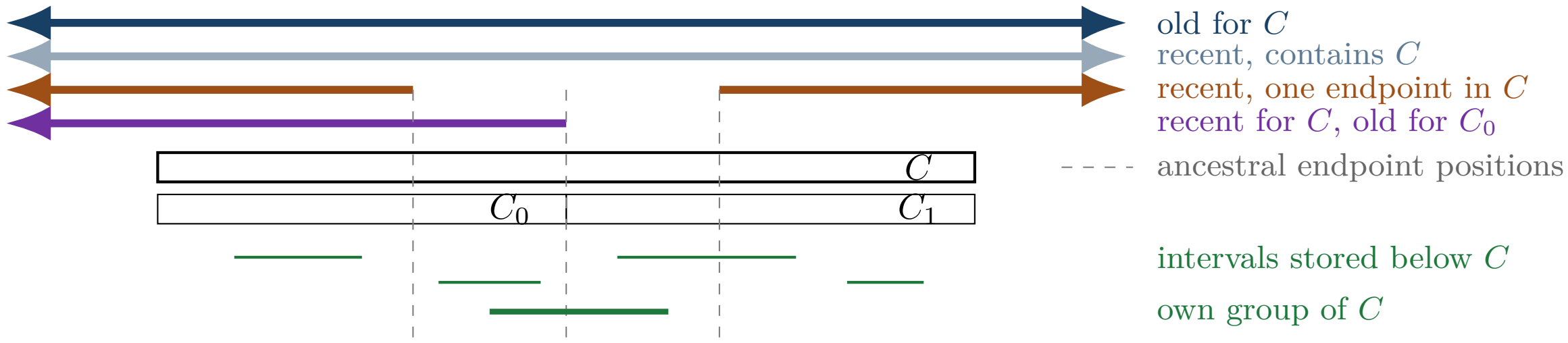

\begin{figure}[t!]
\centering
\begin{tikzpicture}[x=0.2cm,y=1cm,>=Latex,font=\small]
\draw[figwarm,thick,<-] (-7,1.0)--(2,1.0); \fill[figwarm] (2,1.0) circle (1.6pt);
\draw[figwarm,thick,<-] (-7,1.35)--(7,1.35); \fill[figwarm] (7,1.35) circle (1.6pt);
\draw[figwarm,thick,<-] (-7,1.7)--(24,1.7); \fill[figwarm] (24,1.7) circle (1.6pt);
\draw[figwarm,thick,->] (52,1.0)--(71,1.0); \fill[figwarm] (52,1.0) circle (1.6pt);
\fill[figblue!6] (0,0) rectangle (64,0.55);
\foreach \d in {0.5,1,1.5,2,2.5,3,3.5,4,5,6,7,8,10,12,14,16,20,24,28} {
  \draw[figblue,thick] (\d,0.06)--(\d,0.49);
  \draw[figblue,thick] ({64-\d},0.06)--({64-\d},0.49);
}
\draw[figblue,thick] (32,0.06)--(32,0.49);
\draw[thick] (0,0) rectangle (64,0.55);
\node[anchor=east] at (-0.8,0.27) {$C$};
\foreach \d/\t in {4/{\sigma D},8/{2\sigma D},16/{4\sigma D}} {
  \draw[gray,dashed] (\d,-0.05)--(\d,-0.35);
  \node[font=\scriptsize,gray!50!black] at (\d,-0.55) {$\t$};
  \draw[gray,dashed] ({64-\d},-0.05)--({64-\d},-0.35);
}
\draw[gray,dashed] (32,-0.05)--(32,-0.35);
\node[font=\scriptsize,gray!50!black] at (32,-0.55) {$D/2$};
\end{tikzpicture}
\caption{Schematic view of the ancestral endpoint positions in a cell $C$ of length $D$ (\cref{lem:positions}). A rounded interval of a group anchored at a proper ancestor of $C$ that has an endpoint inside $C$ also contains a point outside $C$ (top). An endpoint at distance $d$ from the nearer boundary point of $C$ therefore belongs to an interval of length at least $d$, whose grid spacing is greater than $\sigma d/2$, and every such grid spacing is at least $\max\{1,2\sigma^2D\}$. Hence the possible positions (blue) are dense near the boundary of $C$ and become sparser towards its middle: there are $O(1/\sigma)$ of them at distance less than $\sigma D$, and $O(1/\sigma)$ in each of the $O(\log(1/\sigma))$ ranges $[\nu,2\nu)$ of distances, for $\nu=\sigma D,2\sigma D,4\sigma D,\dots$ (dashed lines).}\label{fig:positions}
\end{figure}
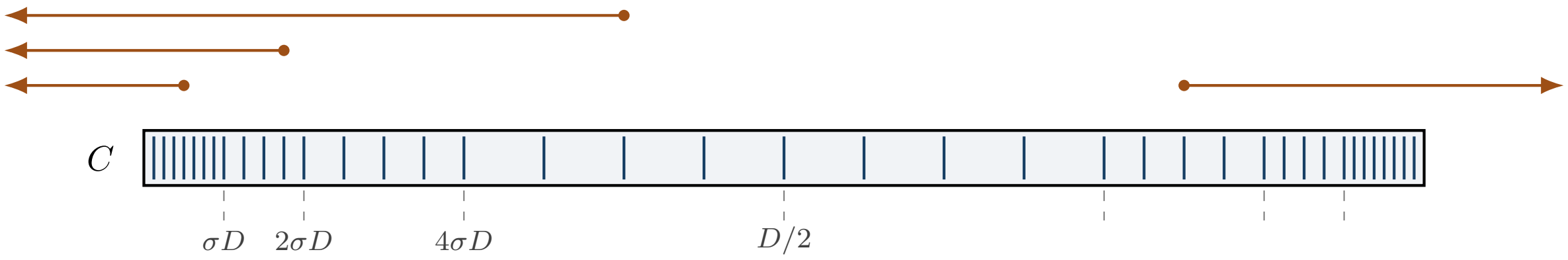

For \cref{thm:structure}, which the dynamic program does not use, it suffices to bound, for every cell $C$, the number of own groups of $C$ and the number of distinct \emph{external neighborhoods} $N_H(v)\setminus V_C$ of the vertices $v\in V_C$: a standard bottom-up construction then produces a clique-width expression in which every vertex built so far carries a label that encodes its external neighborhood (\cref{lem:expression}). Every neighbor of $v$ outside $V_C$ belongs to a recent or an old group of $C$ (\cref{lem:outside-neighbors}), and old groups are adjacent to all of $V_C$. A recent group that meets $C$ but does not contain it has exactly one endpoint inside $C$, and whether $v$ is adjacent to it is decided by comparing one endpoint of $I_v$ with that endpoint. Hence the external neighborhood of $v$ is determined by the positions of the two endpoints of $I_v$ among the \emph{ancestral endpoint positions} of $C$, the endpoints inside $C$ of rounded intervals stored at proper ancestors of $C$ (\cref{fig:cell}). Bounding their number by the number of recent groups would give $O(\sigma^{-4}\log^4(1/\sigma))$ classes; we count the positions directly (\cref{fig:positions}). Such a position at distance $d$ from the nearer boundary point of $C$ belongs to an interval that reaches the midpoint of an ancestor, which lies outside $C$; that interval therefore has length at least $d$, and its grid spacing is at least $\sigma d/2$. If $C$ has length $D$, every proper ancestor has length at least $2D$, so by \cref{sec:ov-delay} all positions lie on the grid of spacing $\max\{1,2\sigma^2D\}$. This leaves $O(1/\sigma)$ positions at distance $d<\sigma D$, and grouping the remaining positions by their distance $d\in[\nu,2\nu)$, where $\nu$ ranges over powers of two, gives $O(1/\sigma)$ positions in each of $O(\log(1/\sigma))$ ranges. So there are $O(\sigma^{-1}\log(1/\sigma))$ ancestral endpoint positions and $O(\sigma^{-2}\log^2(1/\sigma))$ classes, which gives the width bound of \cref{thm:structure}.

\subsection{Hardness on chordal graphs}\label{sec:ov-hardness}

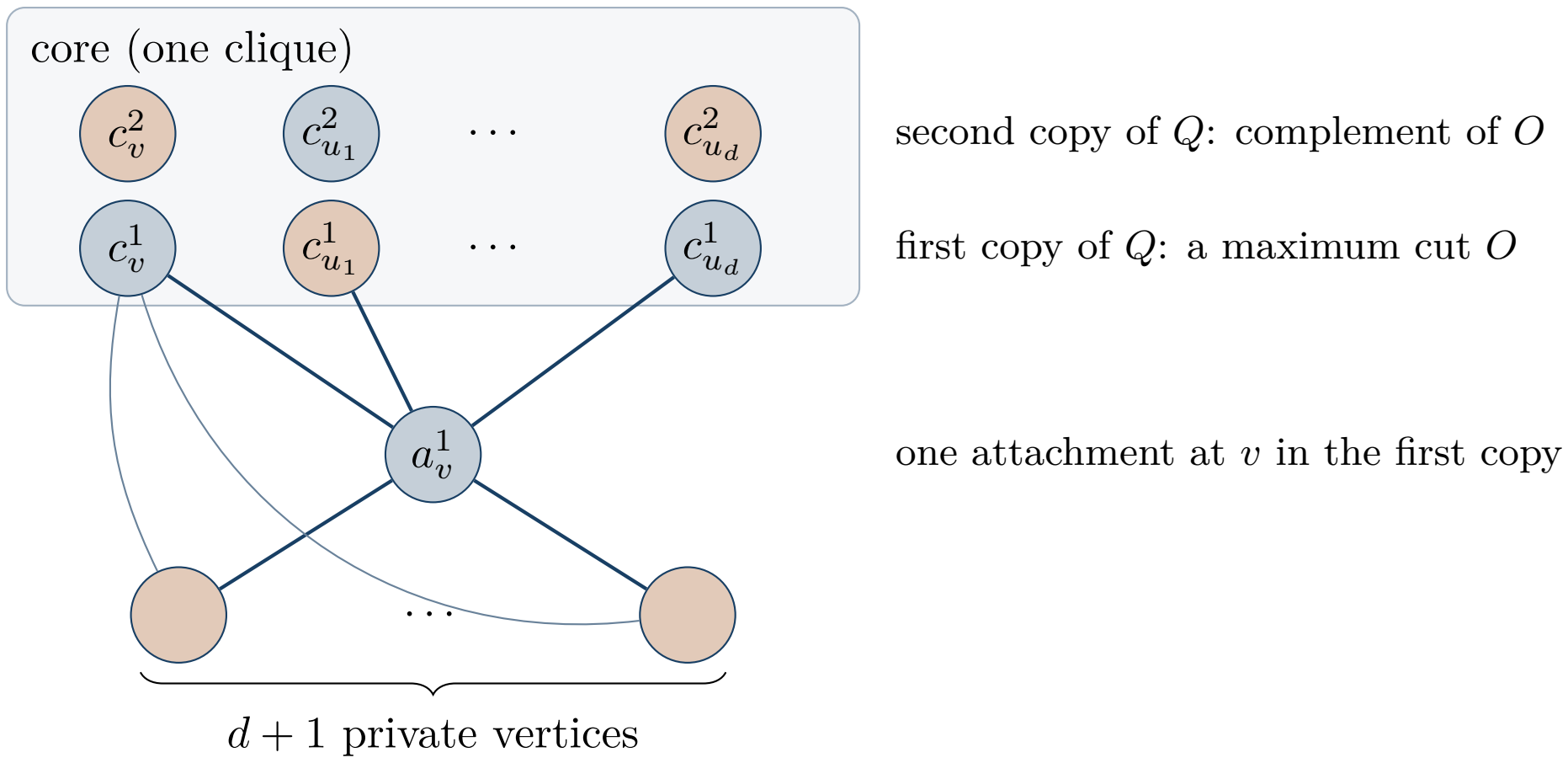
\begin{figure}[t]
\centering
\begin{tikzpicture}[x=1cm,y=0.9cm,>=Latex,
  vtx/.style={circle,draw=figblue,minimum size=7.5mm,inner sep=0.5pt,font=\small},
  sA/.style={fill=figblue!25},
  sB/.style={fill=figwarm!30},
  lab/.style={font=\small,align=center}]
 \draw[rounded corners,fill=figblue!4,draw=figblue!40] (-3.35,0.95) rectangle (3.35,3.55);
 \node[lab,anchor=north west] at (-3.3,3.52) {core (one clique)};
 \node[vtx,sB] (c2v) at (-2.4,2.45) {$c^2_v$};
 \node[vtx,sA] (c2u1) at (-0.8,2.45) {$c^2_{u_1}$};
 \node[lab] at (0.5,2.45) {$\cdots$};
 \node[vtx,sB] (c2ud) at (2.2,2.45) {$c^2_{u_d}$};
 \node[vtx,sA] (cv) at (-2.4,1.45) {$c^1_v$};
 \node[vtx,sB] (u1) at (-0.8,1.45) {$c^1_{u_1}$};
 \node[lab] at (0.5,1.45) {$\cdots$};
 \node[vtx,sA] (ud) at (2.2,1.45) {$c^1_{u_d}$};
 \node[lab,anchor=west,font=\footnotesize] at (3.5,2.45) {second copy of $Q$: complement of $O$};
 \node[lab,anchor=west,font=\footnotesize] at (3.5,1.45) {first copy of $Q$: a maximum cut $O$};
 \node[vtx,sA] (a) at (0,-0.35) {$a^1_v$};
 \draw[figblue,thick] (cv)--(a) (a)--(u1) (a)--(ud);
 \node[vtx,sB] (b1) at (-2,-1.75) {};
 \node[lab] at (0,-1.75) {$\cdots$};
 \node[vtx,sB] (bt) at (2,-1.75) {};
 \draw[figblue,thick] (a)--(b1) (a)--(bt);
 \draw[figblue!65] (cv) to[bend right=18] (b1);
 \draw[figblue!65] (cv) to[bend right=40] (bt);
 \draw[decorate,decoration={brace,amplitude=5pt,mirror}] (-2.3,-2.25)--(2.3,-2.25) node[midway,below=6pt,font=\small]{$d+1$ private vertices};
 \node[lab,anchor=west,font=\footnotesize] at (3.5,-0.35) {one attachment at $v$ in the first copy};
\end{tikzpicture}
\caption{The target graph near a source vertex $v$ with neighbors $u_1,\dots,u_d$ in $Q$. The core is one clique on the core vertices of both copies of $Q$; its edges are not drawn. The middle vertex $a^1_v$ of an attachment at $v$ in the first copy is adjacent to $c^1_v$ and to $c^1_{u_1},\dots,c^1_{u_d}$, and each of its $d+1$ private vertices is adjacent to exactly $c^1_v$ and $a^1_v$; there are $p$ such attachments at every source vertex of each copy. The fill colors show the two sides of an optimal cut: the first copy follows a maximum cut $O$ of $Q$ and the second copy its complement, so that the core is split evenly, and in every attachment the middle vertex lies on the side of its core vertex and the private vertices lie on the other side.}\label{fig:gadget}
\end{figure}

The reduction behind \cref{thm:hardness} starts from Max-Cut on simple graphs, which is NP-hard to approximate within $16/17+\eta$ for every $\eta>0$~\cite{hastad2001,tssw2000,cst2001}. Replacing the vertices of a source graph $Q$, with $n_Q$ vertices and $m_Q$ edges, by the vertices of a clique, the \emph{core}, gives a chordal graph but erases the cut structure of $Q$. We recover it with an attachment at every source vertex $v$, shown in \cref{fig:gadget}: a \emph{middle vertex} $a_v$, adjacent to the core vertex $c_v$ of $v$ and to the core vertices $c_u$ of all neighbors $u$ of $v$, and $\deg_Q(v)+1$ \emph{private vertices}, each adjacent to exactly $c_v$ and $a_v$. Fix the sides of all core vertices, and let $\rho_v$ be the number of neighbors $u$ of $v$ whose vertex $c_u$ lies on the other side than $c_v$. Placing $a_v$ on the side of $c_v$ and the private vertices on the other side cuts $2(\deg_Q(v)+1)+\rho_v$ edges of the attachment, and placing $a_v$ on the other side cuts at most $2(\deg_Q(v)+1)-\rho_v$; so the first choice is optimal, possibly tied when $\rho_v=0$. Summed over $v$, the attachments reward each cut source edge twice, on top of a baseline that does not depend on the cut. The private vertices are needed for this: without them, the best placement of $a_v$ would cut $\max\{\rho_v,\deg_Q(v)+1-\rho_v\}$ edges, which is not an increasing function of $\rho_v$.

A clique with $\kappa$ vertices on one side of a cut and $\kappa'$ on the other contributes $\kappa\kappa'$ cut edges. For a fixed number of clique vertices this term is maximized by an even split, but the side sizes are dictated by the source cut. We therefore take two copies of $Q$ and put the core vertices of both copies into a single clique. An optimal source cut in one copy and its complement in the other attain the source optimum in both copies while splitting the clique evenly, so the target optimum is known exactly. We also repeat every attachment $p$ times, so that the attachments outweigh the clique. The target optimum is then $4p\,(2m_Q+n_Q+\mc(Q))+n_Q^2$, and every target cut decodes to a source cut whose deficit, the optimum minus the value, is at most a $1/(4p)$ fraction of the target's deficit. In this optimum, only the term $4p\cdot\mc(Q)$ counts cut source edges. In \cref{sec:hardness}, $Q$ is obtained from the original source graph by replacing every vertex by $k$ nonadjacent copies, which multiplies the number of edges by $k^2$ and the number of vertices by $k$, and preserves approximation ratios; for large $k$ and $p$, the terms $4p\,n_Q$ and $n_Q^2$ become negligible, while $8p\,m_Q\le16p\cdot\mc(Q)$ because $\mc(Q)\ge m_Q/2$. A target cut within a factor $1-\delta$ of the optimum therefore yields a source cut within about $1-5\delta$, which turns the source gap $1/17$ into the target gap $1/85$. If the private vertices, then the middle vertices, then the core vertices are removed one at a time, the remaining neighbors of every removed vertex form a clique; this perfect elimination ordering certifies that the target graph is chordal, and deleting the private vertices leaves a split graph.

\section{Preliminaries}\label{sec:prelims}

All graphs are finite, simple, undirected, and unweighted. For a graph $G$ we write $n=|V(G)|$ and $m=|E(G)|$, $N_G(v)$ for the set of neighbors of a vertex $v$, and $\deg_G(v)=|N_G(v)|$. For $S\subseteq V(G)$, $\cut_G(S)$ is the number of edges with exactly one endpoint in $S$, $\mc(G)=\max_S\cut_G(S)$, and $\mb(G)$ is the maximum of $\cut_G(S)$ over all $S$ with $|S|=\lfloor n/2\rfloor$. A set $S$ with $|S|=\lfloor n/2\rfloor$ is a \emph{bisection}. Including every vertex in $S$ independently with probability $1/2$ cuts every edge with probability $1/2$, and a uniformly random set of $\lfloor n/2\rfloor$ vertices cuts every edge with probability $2\lfloor n/2\rfloor\lceil n/2\rceil/(n(n-1))\ge1/2$ when $n\ge2$; hence
\begin{equation}\label{eq:half}
\mc(G)\ge m/2\qquad\text{and}\qquad\mb(G)\ge m/2\quad\text{if }n\ge2 .
\end{equation}
Two vertices are \emph{true twins} if they are adjacent and have the same neighbors apart from each other. Logarithms are binary. All algorithms that we give are deterministic. We use the word-RAM model with words of $\Theta(\log n)$ bits and assume that two endpoints of an input representation can be compared in constant time.

\paragraph{Interval graphs.} An \emph{interval representation} of a graph $G$ assigns to every vertex $v$ a closed interval $[\ell_v,r_v]$ of the real line, $\ell_v\le r_v$, such that distinct vertices $u,v$ are adjacent if and only if $[\ell_u,r_u]\cap[\ell_v,r_v]\ne\emptyset$, that is, $\max\{\ell_u,\ell_v\}\le\min\{r_u,r_v\}$. A graph is an \emph{interval graph} if it has an interval representation.

\paragraph{Chordal and split graphs.} A graph is \emph{chordal} if it contains no induced cycle of length at least four. A \emph{perfect elimination ordering} is an ordering of the vertices in which the neighbors of every vertex that come after it form a clique. A graph with a perfect elimination ordering is chordal, because the first vertex of an induced cycle of length at least four would have two nonadjacent later neighbors. A \emph{split graph} is a graph whose vertex set can be partitioned into a clique and an independent set.

\paragraph{Clique-width.} A $w$-labeled graph is a graph together with a map from its vertices to the labels $\{1,\dots,w\}$. A \emph{$w$-expression} builds a $w$-labeled graph with the following operations: create a single vertex with label $i$; take the disjoint union of two $w$-labeled graphs; for labels $i\ne j$, add all edges between the vertices with label $i$ and those with label $j$; and, for labels $i\ne j$, give label $j$ to every vertex with label $i$. The \emph{clique-width} $\cw(H)$ of a graph $H$ is the smallest $w$ such that some $w$-expression builds a labeled graph whose underlying graph is $H$~\cite{co2000}.

\paragraph{Approximation.} An algorithm approximates Max-Cut within a factor $\beta$ on a class of graphs if, on every graph $G$ of the class, it returns a set $S$ with $\cut_G(S)\ge\beta\cdot\mc(G)$. It is NP-hard to approximate Max-Cut within $\beta$ on a class if a polynomial-time algorithm with this guarantee would imply $\Pclass=\NP$.
\section{The edited graph and its hierarchy}\label{sec:structure}

In this section we fix an interval graph $G$ with a representation and an integer $b\ge4$, and we set
\[
\sigma=2^{-b}\qquad\text{and}\qquad\tau=2b .
\]
We construct a spanning subgraph $H$ of $G$ with at most $28\sigma m$ fewer edges (\cref{lem:deleted}), a hierarchy of cells, and a partition of all vertices except some isolated ones, whose total degree in $G$ is small, into groups of true twins, each stored at a cell. The two properties that the later sections use are that every group acts uniformly on all cells that it meets at least $\tau$ levels below the cell where it is stored (\cref{lem:old-universal}), and that only few groups are stored at the $\tau$ levels above any cell (\cref{lem:transitions}). \Cref{prop:structure-summary} collects the results.

\subsection{Rank coordinates}\label{sec:rank}

Sort the $2n$ endpoints of the representation, placing, among endpoints with equal value, every left endpoint before every right endpoint and breaking the remaining ties arbitrarily, and replace each endpoint by its position in this order. For two vertices $u,v$, possibly equal, the comparison $\ell_u\le r_v$ between a left and a right endpoint keeps its truth value: if $\ell_u<r_v$ or $\ell_u>r_v$, the sorted order preserves the relation, and if $\ell_u=r_v$, the left endpoint is placed first. The intervals of distinct $u$ and $v$ intersect exactly when $\ell_u\le r_v$ and $\ell_v\le r_u$, so the new intervals represent the same graph; and taking $u=v$ shows that $\ell_v<r_v$ after the replacement, since the new endpoints are distinct. From now on we use this representation: the $2n$ endpoints are the distinct integers $1,\dots,2n$, and $\ell_v<r_v$ for every vertex $v$. The \emph{rank length} of $v$ is $L_v=r_v-\ell_v\ge1$.

\begin{lemma}\label{lem:rank}
Every vertex $v$ satisfies $L_v\le2\deg_G(v)+1$. If $L_v\ge2$, then $\deg_G(v)\ge1$ and $L_v\le3\deg_G(v)$. Consequently, $\sum_{v:\,L_v\ge2}L_v\le6m$.
\end{lemma}

\begin{proof}
The $L_v-1$ integers strictly between $\ell_v$ and $r_v$ are endpoints of other vertices. If an endpoint of a vertex $u$ lies strictly between $\ell_v$ and $r_v$, then $\ell_u<r_v$ and $\ell_v<r_u$, so $u$ is a neighbor of $v$. Since every neighbor has two endpoints, $L_v-1\le2\deg_G(v)$. If $L_v\ge2$, some endpoint lies strictly between $\ell_v$ and $r_v$, so $\deg_G(v)\ge1$ and $2\deg_G(v)+1\le3\deg_G(v)$. Summing over $v$ gives $\sum_{v:\,L_v\ge2}L_v\le3\sum_v\deg_G(v)=6m$.
\end{proof}

\subsection{The hierarchy and the choice of a shift}\label{sec:hierarchy}

Let $M$ be the smallest power of two with $M\ge2n+1$. The \emph{hierarchy} is the following rooted binary tree of open intervals of the real line, which we call \emph{cells}. The root is $(-2M,2M)$. Every cell $(\zeta,\zeta+D)$ with $D\ge2$ has the two children $(\zeta,\zeta+D/2)$ and $(\zeta+D/2,\zeta+D)$, and the cells with $D=1$ are leaves. We call $|C|=D$ the \emph{length} of $C=(\zeta,\zeta+D)$ and $\zeta+D/2$ its \emph{midpoint}, and we write $\depth(C)$ for the depth of $C$ in the tree, the root having depth $0$. For $D\le2M$, the cells of length $D$ are exactly the open intervals between consecutive multiples of $D$ in $(-2M,2M)$, and the hierarchy has $8M-1=O(n)$ cells. The \emph{subtree} of a cell consists of the cell and its descendants; two cells are \emph{comparable} if one lies in the subtree of the other, and incomparable cells are disjoint.

For a vertex $v$, let $s_v$ be the largest power of two with $s_v\le L_v$, its \emph{scale}; thus $s_v\le L_v<2s_v$. Its \emph{home length} is $D_v=\min\{4M,\,s_v/\sigma\}$, a power of two. For a \emph{shift} $\theta\in\{0,\dots,2M-1\}$ we consider the shifted intervals $[\ell_v-\theta,r_v-\theta]$, which lie in $[2-2M,\,2M-1]$ and hence in the root. A vertex $v$ with $D_v=4M$ has the root as its \emph{home}. A vertex $v$ with $D_v<4M$ has as its home the cell of length $D_v$ that contains $[\ell_v-\theta,r_v-\theta]$, if such a cell exists; otherwise $v$ is \emph{rejected}. Let $Z_\theta$ be the set of rejected vertices.

\begin{lemma}\label{lem:shift}
There is a shift $\theta\in\{0,\dots,2M-1\}$ with $\sum_{v\in Z_\theta}\deg_G(v)\le4\sigma m$.
\end{lemma}

\begin{proof}
Vertices with $D_v=4M$ are never rejected. Fix a vertex $v$ with $D_v<4M$. Then $D_v=s_v/\sigma\ge16s_v>L_v+1$, and $D_v$ divides $2M$. The vertex $v$ is rejected exactly when some multiple of $D_v$ lies in the integer interval $[\ell_v-\theta,r_v-\theta]$, which contains $L_v+1$ integers; this happens for exactly $L_v+1$ residues of $\theta$ modulo $D_v$. As $\theta$ ranges over $\{0,\dots,2M-1\}$, every residue modulo $D_v$ occurs equally often, so $v$ is rejected for a fraction $(L_v+1)/D_v\le2s_v/D_v=2\sigma$ of the shifts. Averaging over all shifts, $\frac1{2M}\sum_\theta\sum_{v\in Z_\theta}\deg_G(v)\le2\sigma\sum_v\deg_G(v)=4\sigma m$, and some shift attains at most the average.
\end{proof}

From now on we fix a shift $\theta$ as in \cref{lem:shift}, replace every endpoint $x$ by $x-\theta$, and write $Z=Z_\theta$. Thus the endpoints are distinct integers in the root, the quantities $L_v$, $s_v$, and $D_v$ are unchanged, and every vertex $v\notin Z$ has a home that contains $[\ell_v,r_v]$. We call the vertices of $Z$ \emph{rejected} and the others \emph{retained}.

\subsection{Rounding}\label{sec:rounding}

For a retained vertex $v$, let $h_v=\max\{1,\sigma s_v\}$, a power of two, and define its \emph{rounded interval} $I_v=[A_v,B_v]$ by
\[
A_v=h_v\left\lceil\ell_v/h_v\right\rceil\qquad\text{and}\qquad B_v=h_v\left\lfloor r_v/h_v\right\rfloor .
\]
The \emph{edited graph} $H$ has vertex set $V(G)$; two retained vertices $u\ne v$ are adjacent in $H$ if $I_u\cap I_v\neq\emptyset$, and rejected vertices are isolated in $H$. In particular, $H$ is an interval graph: the rounded intervals of the retained vertices, together with pairwise disjoint intervals outside the root for the rejected vertices, represent it.

\begin{lemma}\label{lem:rounding}
Every retained vertex $v$ satisfies the following.
\begin{enumerate}[label=(\alph*)]
\item $\ell_v\le A_v<\ell_v+h_v$ and $r_v-h_v<B_v\le r_v$; if $h_v=1$, then $I_v=[\ell_v,r_v]$.
\item $s_v/2\le B_v-A_v<2s_v$; in particular $A_v<B_v$.
\item $I_v$ is contained in the home of $v$.
\end{enumerate}
\end{lemma}

\begin{proof}
Part (a) follows from the definitions, and from the integrality of the endpoints when $h_v=1$. For (b), if $h_v=1$ then $B_v-A_v=L_v\in[s_v,2s_v)$. Otherwise $h_v=\sigma s_v$, and by (a), $B_v-A_v>L_v-2h_v\ge s_v-2\sigma s_v\ge s_v/2$, since $\sigma\le1/16$; moreover $B_v-A_v\le L_v<2s_v$. Part (c) holds because $I_v\subseteq[\ell_v,r_v]$.
\end{proof}

\begin{lemma}\label{lem:deleted}
$H$ is a spanning subgraph of $G$, and $|E(G)\setminus E(H)|\le28\sigma m$.
\end{lemma}

\begin{proof}
Since $I_v\subseteq[\ell_v,r_v]$ for every retained $v$ and rejected vertices are isolated, $H$ is a subgraph of $G$. The edges of $G$ at rejected vertices number at most $\sum_{v\in Z}\deg_G(v)\le4\sigma m$ by \cref{lem:shift}. Now consider an edge $uv$ of $G$ between retained vertices that is not an edge of $H$, and assume that $h_u\le h_v$, breaking ties by a fixed order of the vertices; we \emph{charge} the edge to $v$. If $h_v=1$, then $h_u=1$, both intervals are unchanged by \cref{lem:rounding}(a), and $uv$ would be an edge of $H$; hence $h_v\ge2$.

We claim that an endpoint of $u$ lies at distance less than $h_v$ from $A_v$ or from $B_v$. The rounded intervals are disjoint. If $B_v<A_u$, the original intervals intersect, so $\ell_u\le r_v$, and since endpoints are distinct, $\ell_u<r_v<B_v+h_v$. On the other hand, $\ell_u>A_u-h_u>B_v-h_u\ge B_v-h_v$. Hence $|\ell_u-B_v|<h_v$. If $B_u<A_v$, then symmetrically $\ell_v<r_u$, so $r_u>\ell_v>A_v-h_v$ and $r_u<B_u+h_u<A_v+h_v$; hence $|r_u-A_v|<h_v$. This proves the claim.

Map every edge $uv$ charged to $v$ to an endpoint of $u$ at distance less than $h_v$ from $A_v$ or $B_v$. This map is injective, because distinct edges charged to $v$ have distinct other endpoints $u$ and distinct vertices have disjoint sets of endpoints. Its image lies in the two open intervals of length $2h_v$ centered at the integers $A_v$ and $B_v$, which contain at most $2(2h_v-1)<4h_v$ integers. Hence fewer than $4h_v$ edges are charged to $v$. A vertex with $h_v\ge2$ has $h_v=\sigma s_v\le\sigma L_v$ and $L_v\ge s_v\ge2/\sigma\ge2$, so by \cref{lem:rank} the number of edges charged to some vertex is less than
\[
\sum_{v:\,h_v\ge2}4h_v\;\le\;4\sigma\sum_{v:\,L_v\ge2}L_v\;\le\;24\sigma m .
\]
Together with the edges at rejected vertices, at most $28\sigma m$ edges are deleted.
\end{proof}

\subsection{Anchors and groups}\label{sec:anchors}

For a retained vertex $v$, start at its home and, as long as the current cell has a child that contains $I_v$, move to that child. The cell where this process stops is the \emph{anchor} $\anc(v)$ of $v$. Since cells are open intervals, a cell contains $I_v$ exactly when it contains both $A_v$ and $B_v$.

\begin{lemma}\label{lem:anchor}
Let $v$ be a retained vertex whose anchor has length $D$.
\begin{enumerate}[label=(\alph*)]
\item $I_v\subseteq\anc(v)$, and $I_v$ contains the midpoint of $\anc(v)$; in particular $D\ge2$.
\item $\sigma D\le s_v<2D$, and $h_v\ge\max\{1,\sigma^2D\}$.
\item If $u$ is a retained vertex and $\anc(u)$ and $\anc(v)$ are incomparable, then $I_u\cap I_v=\emptyset$. Hence every edge of $H$ joins two vertices with comparable anchors.
\end{enumerate}
\end{lemma}

\begin{proof}
(a) The home contains $I_v$ by \cref{lem:rounding}(c), and every step of the descent preserves this. Let $(\zeta,\zeta+D)$ be the anchor. It contains the integer $A_v$, so $D\ge2$. Neither child contains $I_v$, while $\zeta<A_v\le B_v<\zeta+D$; hence $A_v\le\zeta+D/2\le B_v$.

(b) The anchor lies in the subtree of the home, so $D\le D_v\le s_v/\sigma$. By (a) and \cref{lem:rounding}(b), $D>B_v-A_v\ge s_v/2$. Therefore $h_v=\max\{1,\sigma s_v\}\ge\max\{1,\sigma^2D\}$.

(c) Incomparable cells are disjoint, and $I_u\subseteq\anc(u)$, $I_v\subseteq\anc(v)$ by (a).
\end{proof}

Two retained vertices belong to the same \emph{group} if they have the same anchor, the same scale, and the same rounded interval. For a group $g$ we write $\anc(g)$, $s_g$, $h_g$, and $I_g=[A_g,B_g]$ for these common values, and $|g|$ for the number of its vertices. Since adjacency between retained vertices in $H$ depends only on their rounded intervals, every group is a clique of true twins in $H$, and two distinct groups $g,g'$ are either completely joined, if $I_g\cap I_{g'}\ne\emptyset$, or have no edges between them. In the first case we write $g\sim g'$.

\begin{lemma}\label{lem:groups-per-anchor}
Every cell is the anchor of at most $(b+1)(2/\sigma+1)^2$ groups.
\end{lemma}

\begin{proof}
Let $C$ be a cell of length $D$ with midpoint $\mu$. By \cref{lem:anchor}(b), the scale $s$ of a group anchored at $C$ is a power of two in $[\sigma D,D]$, so there are at most $b+1$ choices. Fix $s$ and let $h=\max\{1,\sigma s\}$. By \cref{lem:anchor}(a) and \cref{lem:rounding}(b), the rounded interval $[A,B]$ satisfies $A\le\mu\le B$ and $B-A<2s$, so $A\in(\mu-2s,\mu]$ and $B\in[\mu,\mu+2s)$. Both are multiples of $h$, and an interval of length $2s$ contains at most $2s/h+1\le2/\sigma+1$ multiples of $h$: if $h=\sigma s$, then $2s/h=2/\sigma$; otherwise $h=1>\sigma s$, and $2s<2/\sigma$. Hence there are at most $(2/\sigma+1)^2$ rounded intervals for each scale.
\end{proof}

\subsection{Recent and old groups}\label{sec:recent-old}

For a group $g$ and a cell $C$ in the subtree of $\anc(g)$ other than $\anc(g)$, the \emph{age} of $g$ at $C$ is $\depth(C)-\depth(\anc(g))\ge1$. The group $g$ \emph{meets} $C$ if $I_g\cap C\ne\emptyset$. For every cell $C$ we define:
\begin{itemize}
\item $V_C$, the set of retained vertices whose anchor lies in the subtree of $C$, and $W(C)=|V_C|$;
\item $\cA(C)$, the set of groups anchored at $C$, which we call the \emph{own} groups of $C$;
\item $\cR(C)$, the set of groups that are anchored at a proper ancestor of $C$, have age less than $\tau$ at $C$, and meet $C$; we call them \emph{recent} for $C$;
\item $\cO(C)$, the set of groups that are anchored at a proper ancestor of $C$, have age at least $\tau$ at $C$, and meet $C$; we call them \emph{old} for $C$.
\end{itemize}
By \cref{lem:anchor}(a), every vertex $v\in V_C$ satisfies $I_v\subseteq C$.

\begin{lemma}\label{lem:old-universal}
If $g\in\cO(C)$, then $I_g$ contains the closure of $C$, and every vertex of $g$ is adjacent in $H$ to every vertex of $V_C$.
\end{lemma}

\begin{proof}
Since the age of $g$ at $C$ is at least $\tau=2b$, we have $|C|\le2^{-2b}|\anc(g)|=\sigma^2|\anc(g)|$, and $|C|\ge1$; by \cref{lem:anchor}(b), $h_g\ge\sigma^2|\anc(g)|\ge|C|$. Both $h_g$ and $|C|$ are powers of two, so the multiples of $h_g$ are multiples of $|C|$. The endpoints $A_g$ and $B_g$ are multiples of $h_g$, whereas $C$, an open interval between two consecutive multiples of $|C|$, contains no multiple of $|C|$. Hence neither endpoint of $I_g$ lies in $C$. Since $I_g$ meets $C$, it follows that $I_g$ contains the closure of $C$. Every $v\in V_C$ has $I_v\subseteq C\subseteq I_g$, so $v$ is adjacent to every vertex of $g$.
\end{proof}

\begin{lemma}\label{lem:outside-neighbors}
Let $C$ be a cell and $v\in V_C$. Every neighbor of $v$ in $H$ that is not in $V_C$ belongs to a group in $\cR(C)\cup\cO(C)$.
\end{lemma}

\begin{proof}
Let $u\notin V_C$ be a neighbor of $v$ in $H$. Then $u$ is retained, since rejected vertices are isolated, and $\anc(u)$ is comparable with $\anc(v)$ by \cref{lem:anchor}(c). The cells comparable with $\anc(v)$ that lie outside the subtree of $C$ are the proper ancestors of $C$, so $\anc(u)$ is a proper ancestor of $C$. Moreover $I_u$ meets $I_v\subseteq C$, so the group of $u$ meets $C$, and it belongs to $\cR(C)$ or to $\cO(C)$ according to its age at $C$.
\end{proof}

\begin{lemma}\label{lem:transitions}
Let $C$ be a cell and $C'$ a child of $C$.
\begin{enumerate}[label=(\alph*)]
\item $\cR(C')\subseteq\cR(C)\cup\cA(C)$.
\item $\cO(C)\subseteq\cO(C')$, and $\cO(C')\setminus\cO(C)\subseteq\cR(C)$.
\item A group in $\cR(C)\cup\cA(C)$ that belongs neither to $\cR(C')$ nor to $\cO(C')$ has no neighbor in $V_{C'}$.
\item $|\cA(C)|+|\cR(C)|\le\tau(b+1)(2/\sigma+1)^2$.
\end{enumerate}
\end{lemma}

\begin{proof}
(a) A group $g\in\cR(C')$ is anchored at a proper ancestor of $C'$, that is, at $C$ or at a proper ancestor of $C$. In the first case $g\in\cA(C)$. In the second case its age at $C$ is one less than its age at $C'$, hence less than $\tau$, and $g$ meets $C\supseteq C'$; so $g\in\cR(C)$.

(b) A group $g\in\cO(C)$ contains the closure of $C$ by \cref{lem:old-universal}, so it meets $C'$, and its age at $C'$ exceeds its age at $C$; hence $g\in\cO(C')$. Now let $g\in\cO(C')\setminus\cO(C)$. Its age at $C'$ is at least $\tau\ge2$, so $g$ is not anchored at $C$; it is anchored at a proper ancestor of $C$, where its age is at least $\tau-1\ge1$, and it meets $C$. As $g\notin\cO(C)$, its age at $C$ is less than $\tau$, so $g\in\cR(C)$.

(c) Such a group is anchored at a proper ancestor of $C'$. If it met $C'$, it would belong to $\cR(C')$ or to $\cO(C')$ according to its age at $C'$. Hence it does not meet $C'$, whereas every $v\in V_{C'}$ has $I_v\subseteq C'$.

(d) Every group in $\cA(C)\cup\cR(C)$ is anchored at $C$ or at one of the $\tau-1$ nearest proper ancestors of $C$. The bound follows from \cref{lem:groups-per-anchor}.
\end{proof}

\subsection{Summary}\label{sec:structure-summary}

\begin{proposition}\label{prop:structure-summary}
Given an interval representation of an $n$-vertex graph $G$ with $m$ edges and an integer $b\ge4$, one can compute in time $O(n^2)$ a set $Z$ of vertices, a spanning subgraph $H$ of $G$ with $|E(G)\setminus E(H)|\le28\cdot2^{-b}m$ in which the vertices of $Z$ are isolated, the hierarchy, the partition of $V(G)\setminus Z$ into groups with their anchors and rounded intervals, and, for every cell $C$, the sets $\cA(C)$ and $\cR(C)$, the number $W(C)$, and, for every child $C'$ of $C$, the set $\cO(C')\setminus\cO(C)$. These objects satisfy \cref{lem:anchor,lem:groups-per-anchor,lem:old-universal,lem:outside-neighbors,lem:transitions}.
\end{proposition}

\begin{proof}
The correctness statements are \cref{lem:deleted,lem:anchor,lem:groups-per-anchor,lem:old-universal,lem:outside-neighbors,lem:transitions}. For the running time, sorting the endpoints takes $O(n\log n)$ time, the degrees in $G$ can be computed by testing all pairs of intervals, and for each of the $2M=O(n)$ shifts the total degree of the rejected vertices can be computed by testing every vertex; together this takes $O(n^2)$ time. Rounding takes $O(n)$ time, and computing an anchor takes $O(\log n)$ time per vertex. Groups are obtained by sorting the retained vertices by anchor, scale, and rounded interval. Finally, for each of the $O(n)$ cells $C$ and each of the at most $n$ groups $g$, we decide in constant time whether $\anc(g)$ is $C$ or a proper ancestor of $C$, and in the latter case the age of $g$ at $C$ and whether $g$ meets $C$; this classifies $g$ as own, recent, old, or irrelevant for $C$ and yields all listed sets in $O(n^2)$ time. The numbers $W(C)$ are obtained bottom-up, as the sum of the sizes of the own groups of $C$ and the numbers $W$ of its children.
\end{proof}
\section{The approximation scheme}\label{sec:algorithm}

We now prove \cref{thm:eptas}. The algorithm computes the structure of \cref{sec:structure} for a suitable $b$ and then solves Max-Cut and Max-Bisection on the edited graph $H$ approximately, by a dynamic program over the hierarchy. The dynamic program is exact once the number of chosen vertices in each group is restricted to a few allowed values (\cref{lem:recurrence}). Throughout this section we fix $b\ge4$, the structure of \cref{prop:structure-summary}, and a parameter $\gamma\in(0,1)$, and we call the vertices of a set $S\subseteq V(G)$ \emph{red}.

\subsection{Allowed counts}\label{sec:allowed}

Since every group is a clique of true twins in $H$, the number of edges of $H$ cut by $S$ depends only on the numbers $x_g=|S\cap g|$:
\begin{equation}\label{eq:group-objective}
\cut_H(S)=\sum_g x_g(|g|-x_g)+\sum_{\{g,g'\}:\,g\sim g'}\bigl(x_g(|g'|-x_{g'})+(|g|-x_g)x_{g'}\bigr),
\end{equation}
where the second sum ranges over unordered pairs of adjacent groups. For a group $g$, let $\delta_g=\max\{1,\lfloor\gamma|g|\rfloor\}$, and let the \emph{allowed counts} of $g$ be
\[
\cD_g=\{|g|\}\cup\{j\delta_g:\ j\in\mathbb Z,\ 0\le j\delta_g\le|g|\}.
\]
Then $|\cD_g|\le\lfloor2/\gamma\rfloor+2$: if $\delta_g=1$, then $\gamma|g|<2$ and $|\cD_g|=|g|+1$, and otherwise $|g|/\delta_g\le|g|/(\gamma|g|/2)=2/\gamma$. For every integer $0\le x\le|g|$, the largest multiple of $\delta_g$ that is at most $x$ is an allowed count within $\delta_g-1<\max\{1,\gamma|g|\}$ of $x$, and it equals $x$ when $\delta_g=1$; so every count lies within $\gamma|g|$ of an allowed count. A set $S\subseteq V(G)\setminus Z$ is \emph{allowed} if $|S\cap g|\in\cD_g$ for every group $g$.

\begin{lemma}\label{lem:count-rounding}
For every $S\subseteq V(G)$ there is a set $S'$ with $S'\cap Z=S\cap Z$ such that $S'\setminus Z$ is allowed, $\bigl||S'|-|S|\bigr|\le\gamma n$, and $\cut_H(S')\ge\cut_H(S)-2\gamma|E(H)|$.
\end{lemma}

\begin{proof}
For every group $g$, replace $x_g=|S\cap g|$ by the largest allowed count $x'_g\le x_g$, by removing $x_g-x'_g\le\gamma|g|$ vertices of $g$ from $S$. All vertices of $g$ have the same degree $\deg_H(g)$ in $H$, and removing a vertex from $S$ changes the cut by at most its degree. Hence the cut decreases by at most $\sum_g\gamma|g|\deg_H(g)=\gamma\sum_{v}\deg_H(v)=2\gamma|E(H)|$, and $|S|$ decreases by at most $\sum_g\gamma|g|\le\gamma n$.
\end{proof}

\subsection{The tables}\label{sec:tables}

For a cell $C$, a \emph{boundary vector} is a vector $x=(x_g)_{g\in\cR(C)}$ with $x_g\in\cD_g$, which prescribes the number of red vertices in each recent group. For an allowed set $S\subseteq V_C$ (that is, $|S\cap g|\in\cD_g$ for every group $g$ anchored in the subtree of $C$) and a boundary vector $x$, let $\val_C(S;x)$ be the number of edges of $H$ with both endpoints in $V_C$ that are cut by $S$, plus, for every $g\in\cR(C)$ and every $v\in V_C$ adjacent to $g$, the number $|g|-x_g$ if $v\in S$ and the number $x_g$ if $v\notin S$. In words, $\val_C(S;x)$ counts the cut edges inside the subtree of $C$ and between the subtree and the recent groups, assuming that $x_g$ vertices of each recent group $g$ are red; edges to old groups are not counted. For $t\in\{0,\dots,W(C)\}$ let
\[
F_C(x,t)=\max\{\val_C(S;x):\ S\subseteq V_C\text{ allowed},\ |S|=t\},
\]
with $F_C(x,t)=-\infty$ if no such $S$ exists. At the root, $\cR=\cO=\emptyset$ and $V_{\text{root}}=V(G)\setminus Z$, so $F_{\text{root}}(t)$ is the largest number of edges of $H$ cut by an allowed set of size $t$.

To state the recurrence, fix a cell $C$, a boundary vector $x$ for $C$, and a vector $y=(y_g)_{g\in\cA(C)}$ with $y_g\in\cD_g$ for the own groups, and write $|y|=\sum_gy_g$. Let
\begin{align*}
\Loc_C(x,y)=&\sum_{g\in\cA(C)}y_g(|g|-y_g)+\sum_{\substack{\{g,g'\}\subseteq\cA(C)\\ g\sim g'}}\bigl(y_g(|g'|-y_{g'})+(|g|-y_g)y_{g'}\bigr)\\
&+\sum_{\substack{g\in\cA(C),\,g'\in\cR(C)\\ g\sim g'}}\bigl(y_g(|g'|-x_{g'})+(|g|-y_g)x_{g'}\bigr)
\end{align*}
be the number of cut edges inside and between own groups and between own and recent groups. Let $C_0$ and $C_1$ be the children of $C$; if $C$ is a leaf, the terms for the children below are absent. For $j\in\{0,1\}$, let $x^{(j)}$ be the boundary vector for $C_j$ that takes the value $x_g$ for $g\in\cR(C)$ and $y_g$ for $g\in\cA(C)$; it is well defined by \cref{lem:transitions}(a). Let
\[
\Delta U_j=\sum_{g\in\cO(C_j)\setminus\cO(C)}|g|\qquad\text{and}\qquad\Delta R_j=\sum_{g\in\cO(C_j)\setminus\cO(C)}x_g ,
\]
which are well defined by \cref{lem:transitions}(b); they are the number of vertices, and of red vertices, in the groups that are recent for $C$ and old for $C_j$.

\begin{lemma}\label{lem:recurrence}
For every cell $C$, boundary vector $x$, and $t\in\{0,\dots,W(C)\}$,
\[
F_C(x,t)=\max_{y,\,t_0,\,t_1}\Bigl\{\Loc_C(x,y)+\sum_{j\in\{0,1\}}\Bigl(F_{C_j}\bigl(x^{(j)},t_j\bigr)+t_j(\Delta U_j-\Delta R_j)+\bigl(W(C_j)-t_j\bigr)\Delta R_j\Bigr)\Bigr\},
\]
where the maximum ranges over vectors $y$ of allowed counts for the own groups of $C$ and integers $0\le t_j\le W(C_j)$ with $|y|+t_0+t_1=t$.
\end{lemma}

\begin{proof}
The set $V_C$ is the disjoint union of the vertices of the own groups of $C$ and the sets $V_{C_0}$ and $V_{C_1}$. Let $S\subseteq V_C$ be allowed, and let $y_g=|S\cap g|$ for $g\in\cA(C)$ and $S_j=S\cap V_{C_j}$. We sort the edges counted by $\val_C(S;x)$ by the location of their endpoints. Here, an edge between a vertex $v\in V_C$ and a recent group $g$ counts as cut with the multiplicities given in the definition of $\val_C$.
\begin{enumerate}[label=(\roman*)]
\item Edges inside and between own groups, and between own and recent groups of $C$: they contribute $\Loc_C(x,y)$.
\item Edges inside $V_{C_j}$, and between $V_{C_j}$ and the groups of $\cR(C_j)$: they contribute $\val_{C_j}(S_j;x^{(j)})$. Indeed, by \cref{lem:transitions}(a) every group in $\cR(C_j)$ is an own group of $C$, which has $y_g$ red vertices, or a recent group of $C$, for which $x$ prescribes $x_g$.
\item Edges between $V_{C_j}$ and the groups of $\cO(C_j)\setminus\cO(C)$: by \cref{lem:old-universal}, every vertex of these groups is adjacent to every vertex of $V_{C_j}$, so they contribute $|S_j|(\Delta U_j-\Delta R_j)+(W(C_j)-|S_j|)\Delta R_j$.
\end{enumerate}
These classes are disjoint, and they contain every edge counted by $\val_C(S;x)$. There are no edges between $V_{C_0}$ and $V_{C_1}$, by \cref{lem:anchor}(c). An own or recent group of $C$ that is adjacent to a vertex of $V_{C_j}$ belongs to $\cR(C_j)$ or to $\cO(C_j)\setminus\cO(C)$ by \cref{lem:transitions}(c), and then the edge is in class~(ii) or~(iii); an own group of $C$ has age $1<\tau$ at $C_j$, so only the first alternative applies to it; the groups of $\cO(C)$ belong to $\cO(C_j)$ by \cref{lem:transitions}(b), and their edges are counted neither by $\val_C$ nor by any of the three classes. Therefore
\[
\val_C(S;x)=\Loc_C(x,y)+\sum_{j}\Bigl(\val_{C_j}(S_j;x^{(j)})+|S_j|(\Delta U_j-\Delta R_j)+(W(C_j)-|S_j|)\Delta R_j\Bigr).
\]
Conversely, for every choice of allowed counts $y$ and of allowed sets $S_j\subseteq V_{C_j}$, choosing $y_g$ vertices from each own group $g$ and adding $S_0\cup S_1$ gives an allowed set $S\subseteq V_C$ for which this identity holds. Maximizing over $S$ with $|S|=t$ is thus the same as maximizing over $y$ and over $S_0,S_1$ with $|y|+|S_0|+|S_1|=t$, and for fixed $y$ and fixed sizes $t_j=|S_j|$ the two children can be optimized independently. This gives the recurrence.
\end{proof}

Over the whole recursion, every edge of $H$ is counted exactly once. Let $uv$ be an edge of $H$ such that $\anc(u)$ is $\anc(v)$ or a proper ancestor of it (\cref{lem:anchor}(c)), and let $g$ be the group of $u$. If $g$ is an own group of $\anc(v)$ or recent for $\anc(v)$, the edge is counted in $\Loc_{\anc(v)}$. Otherwise $g$ is old for $\anc(v)$; on the path from $\anc(g)$ to $\anc(v)$, let $C'$ be the cell at which $g$ has age exactly $\tau$, with parent $C$. Then $g\in\cR(C)$ and $g\in\cO(C')\setminus\cO(C)$, and the edge is counted by the affine correction for the child $C'$ in the recurrence at $C$, that is, when the dynamic program passes from $C$ to $C'$, and nowhere else.

\begin{lemma}\label{lem:dp-time}
All tables $F_C$ can be computed, together with an allowed set attaining each finite entry of the root table, in time $\omega^{\Lambda}\cdot\Lambda^{O(1)}\cdot O(n^2)$, where $\omega=\lfloor2/\gamma\rfloor+2$ and $\Lambda=\tau(b+1)(2/\sigma+1)^2$.
\end{lemma}

\begin{proof}
We compute the tables bottom-up. We store the table $F_C$ as an array indexed by $t$ and by the boundary vectors $x$, each encoded by the positions of its entries in the sets $\cD_g$; an entry is located with $\Lambda^{O(1)}$ word operations. With each entry we store a maximizing triple $(y,t_0,t_1)$. Since $\sum_C(W(C)+1)=O(n\log n)$, as shown below, all tables together take space $\omega^\Lambda\cdot\Lambda^{O(1)}\cdot O(n\log n)$. At a cell $C$ there are at most $\omega^{|\cR(C)|+|\cA(C)|}\le\omega^\Lambda$ pairs $(x,y)$ by \cref{lem:transitions}(d). For each pair we compute $\Loc_C(x,y)$, the vectors $x^{(j)}$, and the numbers $\Delta U_j,\Delta R_j$ in $\Lambda^{O(1)}$ time, using the sets provided by \cref{prop:structure-summary}. Then, for all $t_0,t_1$ at once, we combine the two rows $F_{C_j}(x^{(j)},\cdot)$, each shifted by its affine correction, by a $(\max,+)$-convolution in $O((W(C_0)+1)(W(C_1)+1))$ time, and update the entries $F_C(x,|y|+t_0+t_1)$, recording a maximizing choice for backtracking. Every pair of distinct vertices contributes to the product $W(C_0)W(C_1)$ only at the cell where the paths from the root to their anchors separate, so $\sum_CW(C_0)W(C_1)\le\binom n2$. Every vertex lies in $V_C$ for at most $\log(4M)+1=O(\log n)$ cells, so $\sum_C(W(C_0)+W(C_1)+1)=O(n\log n)$. Hence the total time is $\omega^\Lambda\cdot\Lambda^{O(1)}\cdot O(n^2)$. Following the recorded choices from a finite entry of the root table visits every cell once, in $\Lambda^{O(1)}$ time per cell, and determines the counts of all groups; choosing that many vertices from each group gives an allowed set attaining the entry.
\end{proof}

\subsection{Proof of the main theorem}\label{sec:eptas-proof}

\begin{restatedtheorem}{thm:eptas}
\thmEptasText
\end{restatedtheorem}

\begin{proof}
If $n<2$ or $m=0$, every set is a maximum cut and every set with $\lfloor n/2\rfloor$ vertices is a maximum bisection. Otherwise let
\[
b=\bigl\lceil\log(448/\eps)\bigr\rceil\qquad\text{and}\qquad\gamma=\eps/16 .
\]
Then $b\ge9$ and $28\sigma=28\cdot2^{-b}\le\eps/16$. We compute the structure of \cref{prop:structure-summary} in $O(n^2)$ time and all tables by \cref{lem:dp-time}. Since $1/\sigma<896/\eps$, $b=O(\log(2/\eps))$, and $\omega=O(1/\eps)$, we have $\Lambda=O(\eps^{-2}\log^2(2/\eps))$ and $\omega^\Lambda\cdot\Lambda^{O(1)}=2^{O(\eps^{-2}\log^3(2/\eps))}$, which gives the claimed running time. Recall that $\cut_G(S)\ge\cut_H(S)$ for all $S$, because $H\subseteq G$, and that $|E(G)\setminus E(H)|\le28\sigma m\le\eps m/16$ by \cref{lem:deleted}.

\emph{Max-Cut.} The algorithm outputs an allowed set $S$ attaining $\max_tF_{\text{root}}(t)$; the vertices of $Z$ are not added. Let $S^*$ be a maximum cut of $G$. Then $\cut_H(S^*)\ge\mc(G)-\eps m/16$, and \cref{lem:count-rounding} gives a set $S'$ with $\cut_H(S')\ge\cut_H(S^*)-2\gamma m$ such that $S'\setminus Z$ is allowed. Since the vertices of $Z$ are isolated in $H$,
\begin{align*}
\cut_G(S)\ge\cut_H(S)&\ge F_{\text{root}}\bigl(|S'\setminus Z|\bigr)\ge\cut_H(S')\\
&\ge\mc(G)-\tfrac{\eps}{16}m-\tfrac{\eps}8m\ge\mc(G)-\tfrac{\eps}{2}m\ge(1-\eps)\mc(G),
\end{align*}
where the last step uses $\mc(G)\ge m/2$ from~\eqref{eq:half}.

\emph{Max-Bisection.} Let $k=\lfloor n/2\rfloor$. Call an integer $t\in\{0,\dots,W(\text{root})\}$ \emph{eligible} if some $z\in\{0,\dots,|Z|\}$ satisfies $|t+z-k|\le\gamma n$. The algorithm chooses, among the eligible $t$ with $F_{\text{root}}(t)>-\infty$, one, $t^*$, that maximizes $F_{\text{root}}(t^*)$; such $t$ exist, as shown below. It then chooses an allowed set $S_1$ with $|S_1|=t^*$ attaining it, and $z^*\in\{0,\dots,|Z|\}$ that minimizes $d=|t^*+z^*-k|$; it adds $z^*$ vertices of $Z$ to $S_1$. Then $d\le\gamma n$. If $|S_1|>k$, it removes from $S_1$ the $d$ vertices of $S_1$ with the smallest degrees in $G$; if $|S_1|<k$, it adds to $S_1$ the $d$ vertices outside $S_1$ with the smallest degrees in $G$. The result $S_2$ has $|S_2|=k$.

The class from which the $d$ vertices are taken has at least $n/2$ vertices: if $|S_1|>k$ then $|S_1|\ge k+1\ge n/2$, and if $|S_1|<k$ then $|V\setminus S_1|>n-k\ge n/2$. Hence the moved vertices have total degree at most $\frac{d}{n/2}\cdot2m\le4\gamma m$. Every edge that is cut by $S_1$ but not by $S_2$ has a moved endpoint, so $\cut_G(S_2)\ge\cut_G(S_1)-4\gamma m$.

Let $S^*$ be a maximum bisection of $G$, and let $S'$ be the set given by \cref{lem:count-rounding}. Then $|S'|$ differs from $k$ by at most $\gamma n$ and $S'\cap Z=S^*\cap Z$, so $t'=|S'\setminus Z|$ is eligible with $z=|S'\cap Z|$. Therefore
\begin{align*}
\cut_G(S_2)&\ge\cut_G(S_1)-4\gamma m\ge\cut_H(S_1)-4\gamma m=F_{\text{root}}(t^*)-4\gamma m\ge F_{\text{root}}(t')-4\gamma m\\
&\ge\cut_H(S')-4\gamma m\ge\cut_H(S^*)-6\gamma m\ge\mb(G)-\tfrac{\eps}{16}m-\tfrac{6\eps}{16}m\ge(1-\eps)\mb(G),
\end{align*}
where we used that adding vertices of $Z$ does not change $\cut_H$, and $\mb(G)\ge m/2$ from~\eqref{eq:half}. Choosing $t^*$, $z^*$, and the moved vertices takes $O(n\log n)$ time.
\end{proof}
\section{Clique-width of the edited graph}\label{sec:cliquewidth}

We now prove \cref{thm:structure}. By the standard argument recalled in \cref{lem:expression}, it suffices to bound, for every cell $C$, the number of distinct neighborhoods that the vertices of $V_C$ have outside $V_C$. Throughout this section we fix $b\ge4$ and the structure of \cref{prop:structure-summary}. For a cell $C$ and $v\in V_C$, the \emph{external neighborhood} of $v$ is $N_H(v)\setminus V_C$, and $K(C)$ denotes the number of distinct external neighborhoods of the vertices of $V_C$.

\subsection{Ancestral endpoints inside a cell}\label{sec:positions}

For a cell $C$ other than the root, let $P(C)$ be the set of integers in $C$ that are an endpoint $A_g$ or $B_g$ of a group $g$ anchored at a proper ancestor of $C$. For a point $\xi\in C$, let $d_C(\xi)$ be the distance from $\xi$ to the nearer boundary point of $C$.

\begin{lemma}\label{lem:positions}
Let $C$ be a cell of length $D$ other than the root, let $g$ be a group anchored at a proper ancestor of $C$, and let $\xi\in P(C)$ be an endpoint of $I_g$. Then $h_g\ge\max\{1,2\sigma^2D\}$ and $h_g>\sigma d_C(\xi)/2$. Consequently, $|P(C)|\le10b/\sigma$.
\end{lemma}

\begin{proof}
The anchor of $g$ is a proper ancestor of $C$ and has length at least $2D$, so $h_g\ge\max\{1,2\sigma^2D\}$ by \cref{lem:anchor}(b). The interval $I_g$ contains the midpoint of $\anc(g)$, which is a boundary point of a child of $\anc(g)$ and therefore does not lie in $C$. Since $I_g$ contains $\xi\in C$ and a point outside $C$, it contains the segment from $\xi$ to a boundary point of $C$, so $B_g-A_g\ge d_C(\xi)$. By \cref{lem:rounding}(b), $s_g>d_C(\xi)/2$, and hence $h_g\ge\sigma s_g>\sigma d_C(\xi)/2$.

For the bound on $|P(C)|$, note that $d_C(\xi)\le D/2$ for all $\xi\in C$. Let $q_\ast=\max\{1,2\sigma^2D\}$, which is a power of two because $D$ and $\sigma$ are. Every $h_g$ as above is a power of two that is at least $q_\ast$, so all elements of $P(C)$ are multiples of $q_\ast$. The points $\xi\in C$ with $d_C(\xi)<\sigma D$ form two intervals of length $\sigma D$, each containing at most $\sigma D/q_\ast+1\le1/(2\sigma)+1$ multiples of $q_\ast$; here we use $\sigma D/q_\ast=1/(2\sigma)$ if $q_\ast=2\sigma^2D$, and $\sigma D\le1/(2\sigma)$ if $q_\ast=1$. Next, for $i=0,\dots,b-1$ let $\nu_i=2^i\sigma D$, and consider the points $\xi\in P(C)$ with $\nu_i\le d_C(\xi)<2\nu_i$; since $\nu_{b-1}=D/2$, these $b$ ranges together with the first case cover all of $P(C)$. For such $\xi$, the spacing $h_g$ is a power of two with $h_g\ge1$ and $h_g>\sigma\nu_i/2$. Let $q_i$ be the largest power of two with $q_i\le\max\{1,\sigma\nu_i/2\}$. Then $q_i$ divides $h_g$, so $\xi$ is a multiple of $q_i$, and $q_i\ge\max\{1,\sigma\nu_i/4\}$. The points with $\nu_i\le d_C(\xi)<2\nu_i$ form two intervals of length $\nu_i$, each containing at most $\nu_i/q_i+1\le4/\sigma+1$ multiples of $q_i$. Altogether,
\[
|P(C)|\le2\Bigl(\frac1{2\sigma}+1\Bigr)+b\cdot2\Bigl(\frac4\sigma+1\Bigr)=\frac{1+8b}\sigma+2+2b\le\frac{10b}\sigma,
\]
where the last step uses $1/\sigma\ge16$ and $b\ge4$.
\end{proof}

\begin{lemma}\label{lem:classes}
For every cell $C$ other than the root, $K(C)\le\bigl(2|P(C)|+1\bigr)^2\le(20b/\sigma+1)^2$. Moreover, $K(\text{root})\le1$.
\end{lemma}

\begin{proof}
Every vertex outside $V_{\text{root}}$ is rejected and hence isolated, so all vertices of $V_{\text{root}}$ have the empty external neighborhood. Now let $C$ be another cell and $v\in V_C$. By \cref{lem:outside-neighbors}, the external neighborhood of $v$ consists of vertices of groups in $\cR(C)\cup\cO(C)$, and since groups are true twins, it is the union of the groups $g\in\cR(C)\cup\cO(C)$ with $I_g\cap I_v\ne\emptyset$. By \cref{lem:old-universal}, every group of $\cO(C)$ belongs to the external neighborhood of every vertex of $V_C$, and so does every group of $\cR(C)$ whose interval contains the closure of $C$, because $I_v\subseteq C$.

Consider a group $g$ that is anchored at a proper ancestor of $C$, meets $C$, and does not contain the closure of $C$; every group of $\cR(C)$ that does not contain the closure of $C$ is of this kind. We claim that exactly one endpoint of $I_g$ lies in $C$. If both did, then $I_g\subseteq C$, and the descent defining $\anc(g)$ would not have stopped at a proper ancestor of $C$, because the child of $\anc(g)$ that contains $C$ would contain $I_g$. If none did, then, since $I_g$ meets $C$, it would contain the closure of $C$. Say $A_g\in C$ and $B_g$ is at least the right boundary point of $C$; the other case is symmetric. Since $I_v\subseteq C$, we have $A_v<B_g$, so $I_g\cap I_v\ne\emptyset$ holds exactly when $A_g\le B_v$, and $A_g\in P(C)$. In the symmetric case, $I_g\cap I_v\ne\emptyset$ holds exactly when $A_v\le B_g$, and $B_g\in P(C)$.

Hence the external neighborhood of $v$ is determined by the position of $A_v$ and of $B_v$ relative to the elements of $P(C)$, that is, by which element of $P(C)$ each of them equals, or between which two consecutive elements, or before the first or after the last element, it lies. There are $2|P(C)|+1$ such positions for each endpoint, which proves the first bound; the second follows from \cref{lem:positions}.
\end{proof}

\subsection{A clique-width expression}\label{sec:expression}

The following lemma turns the bound on external neighborhoods into a clique-width expression. Its argument is standard: the labels of the expression built for the subtree of a cell represent the external neighborhoods of the vertices built so far, as in the relation between module-width and clique-width~\cite{rao2006}.

\begin{lemma}\label{lem:expression}
Suppose that $K(C)\le K$ and $|\cA(C)|\le J$ for every cell $C$. Then $H$ has a $(3K+J+1)$-expression, and such an expression can be computed in time polynomial in $n$, $K$, and $J$.
\end{lemma}

\begin{proof}
We use three palettes $\Pi_0,\Pi_1,\Pi_2$ of $K$ labels each, a palette $\Pi_3$ of $J$ labels, all disjoint, and one further label $\star$. For every cell $C$ with $V_C\ne\emptyset$, in a bottom-up order, we construct an expression $X_C$ that builds $H[V_C]$ in which every vertex carries a label of $\Pi_2$, and two vertices carry the same label exactly when they have the same external neighborhood with respect to $C$. Consider such a cell $C$ with children $C_0$ and $C_1$.
\begin{enumerate}[label=(\arabic*)]
\item For $j\in\{0,1\}$ with $V_{C_j}\ne\emptyset$, take $X_{C_j}$ and relabel its labels from $\Pi_2$ to distinct labels of $\Pi_j$. Its intermediate operations may use any labels, since only its result matters.
\item For every own group $g\in\cA(C)$ choose a distinct label $\pi_g\in\Pi_3$, and build the clique on $g$ as follows: create one vertex with label $\pi_g$; then, $|g|-1$ times, take the disjoint union with a new vertex with label $\star$, add all edges between the labels $\star$ and $\pi_g$, and relabel $\star$ to $\pi_g$.
\item Take the disjoint union of the graphs from (1) and (2). For every pair of adjacent own groups $g\sim g'$, add all edges between $\pi_g$ and $\pi_{g'}$. For every own group $g$ and every label $\lambda\in\Pi_0\cup\Pi_1$ whose vertices are adjacent to $g$, add all edges between $\pi_g$ and $\lambda$.
\item Relabel every label of $\Pi_0\cup\Pi_1\cup\Pi_3$ to the label of $\Pi_2$ that represents the external neighborhood of its vertices with respect to $C$.
\end{enumerate}
Step~(3) is correct for the following reasons. There are no edges between $V_{C_0}$ and $V_{C_1}$ by \cref{lem:anchor}(c). The vertices with a common label $\lambda\in\Pi_j$ have the same neighbors outside $V_{C_j}$, and the own groups of $C$ lie outside $V_{C_j}$; since every own group consists of true twins, either all or none of the edges between $g$ and the vertices with label $\lambda$ are present in $H$. In step~(4), the external neighborhood of a vertex $v\in V_{C_j}$ with respect to $C$ is its external neighborhood with respect to $C_j$ minus the vertices of $V_C$, so it is determined by the label of $v$ after step~(1); and all vertices of an own group have the same external neighborhood. Since the labels of $\Pi_2$ are not in use before step~(4), relabeling one label at a time never merges two classes that should stay apart. At the root, finally, we add every rejected vertex as an isolated vertex with an arbitrary label. The expression uses $3K+J+1$ labels. Step~(2) uses $O(|g|)$ operations for each group $g$, hence $O(n)$ operations in total, and the other steps use $O((K+J)^2)$ operations for each of the $O(n)$ cells. To compute the classes, we sort, for every cell $C$, the vertices of $V_C$ by their rows of the adjacency matrix of $H$ restricted to the columns of $V(G)\setminus V_C$; by radix sort this takes $O(n\,W(C))$ time, and $O(n^2\log n)$ time over all cells, since $\sum_CW(C)=O(n\log n)$.
\end{proof}

\subsection{Proof of the structural theorem}\label{sec:structure-proof}

\begin{restatedtheorem}{thm:structure}
\thmStructureText
\end{restatedtheorem}

\begin{proof}
Let $b=\lceil\log(28/\eps)\rceil$, so that $b\ge5$. Then $28\sigma\le\eps$, so $H$ is a spanning subgraph of $G$ with $|E(G)\setminus E(H)|\le\eps m$ by \cref{lem:deleted}, and $H$ is an interval graph (\cref{sec:rounding}). Moreover, $1/\sigma<56/\eps$ because $\eps<1$. By \cref{lem:classes}, every cell has $K(C)\le(20b/\sigma+1)^2=O(\eps^{-2}\log^2(2/\eps))$, and by \cref{lem:groups-per-anchor}, $|\cA(C)|\le(b+1)(2/\sigma+1)^2=O(\eps^{-2}\log(2/\eps))$. \Cref{lem:expression} therefore gives an expression of width $O(\eps^{-2}\log^2(2/\eps))$. The structure is computed in $O(n^2)$ time by \cref{prop:structure-summary}, and the expression in time polynomial in $n$ and $1/\eps$ by \cref{lem:expression}.
\end{proof}
\section{Hardness on chordal graphs}\label{sec:hardness}

We prove \cref{thm:hardness} by a gap-preserving reduction from Max-Cut on simple graphs. The reduction places two copies of the source graph in one clique and attaches to every source vertex a gadget that rewards the cut edges at that vertex (\cref{lem:local}); two complementary copies make the optimum of the target graph an exact function of the optimum of the source (\cref{lem:decoder}). This section does not use \cref{sec:structure,sec:algorithm,sec:cliquewidth}.

\subsection{The source problem and amplification}\label{sec:source}

\begin{theorem}[\cite{hastad2001,tssw2000,cst2001}]\label{thm:source}
For every $\eta>0$, it is NP-hard to approximate Max-Cut on simple graphs within a factor of $16/17+\eta$.
\end{theorem}

Crescenzi, Silvestri, and Trevisan~\cite[full version, Corollary~18]{cst2001} prove that, for every $\eta'>0$, unless $\Pclass=\NP$ no polynomial-time algorithm computes for every simple graph $G$ a set $S$ with $\cut_G(S)\ge\mc(G)/(17/16-\eta')$. They derive this from the same hardness for Max-Cut with edge weights, which follows from the work of H{\aa}stad~\cite{hastad2001} with the gadgets of~\cite{tssw2000}, by a reduction from the weighted to the unweighted problem. Given $\eta>0$, choose $\eta'>0$ with $1/(17/16-\eta')\le16/17+\eta$. An algorithm that approximates Max-Cut within $16/17+\eta$ also approximates it within $1/(17/16-\eta')$, which proves \cref{thm:source}.

For a simple graph $Q_0$ and an integer $k\ge1$, let $Q_0^{(k)}$ be the graph obtained by replacing every vertex $v$ of $Q_0$ by $k$ pairwise nonadjacent copies and every edge $uv$ by the $k^2$ edges between the copies of $u$ and the copies of $v$.

\begin{lemma}\label{lem:blowup}
$\mc(Q_0^{(k)})=k^2\mc(Q_0)$. Moreover, from any set $S$ of vertices of $Q_0^{(k)}$, a set $S_0\subseteq V(Q_0)$ with $\cut_{Q_0}(S_0)\ge\cut_{Q_0^{(k)}}(S)/k^2$ can be computed in polynomial time.
\end{lemma}

\begin{proof}
For a set $S$ of vertices of $Q_0^{(k)}$ and $v\in V(Q_0)$, let $\lambda_v$ be the fraction of the copies of $v$ that lie in $S$. Then
\[
\cut_{Q_0^{(k)}}(S)=k^2\,\Phi(\lambda),\qquad\text{where}\quad\Phi(\lambda)=\sum_{uv\in E(Q_0)}\bigl(\lambda_u(1-\lambda_v)+(1-\lambda_u)\lambda_v\bigr).
\]
Since $Q_0$ has no loops, $\Phi$ is an affine function of each coordinate $\lambda_v$ when the others are fixed. Processing the vertices of $Q_0$ one by one and replacing $\lambda_v$ by whichever of $0$ and $1$ gives the larger value of $\Phi$ never decreases $\Phi$, and it ends with the indicator vector of a set $S_0$ with $\cut_{Q_0}(S_0)=\Phi\ge\cut_{Q_0^{(k)}}(S)/k^2$. Each step evaluates $\Phi$ twice on a vector of rationals with denominator $k$, so $S_0$ is computed in polynomial time. Conversely, taking all copies of the vertices of a maximum cut of $Q_0$ gives a cut of $Q_0^{(k)}$ of size $k^2\mc(Q_0)$.
\end{proof}

\subsection{The reduction}\label{sec:reduction}

Let $Q$ be a simple graph with $n_Q$ vertices and $m_Q\ge1$ edges, and let $p\ge1$ be an integer. The graph $T_p(Q)$ has the following vertices and edges.
\begin{itemize}
\item \emph{Core vertices} $c^i_v$ for $i\in\{1,2\}$ and $v\in V(Q)$; all $2n_Q$ core vertices are pairwise adjacent.
\item For every $i\in\{1,2\}$, $v\in V(Q)$, and $j\in\{1,\dots,p\}$, an \emph{attachment} consisting of a \emph{middle vertex} $a^{i,j}_v$, adjacent to $c^i_v$ and to $c^i_u$ for every neighbor $u$ of $v$ in $Q$, and of $\deg_Q(v)+1$ \emph{private vertices}, each adjacent to exactly $c^i_v$ and $a^{i,j}_v$; we call it the attachment $(i,v,j)$.
\end{itemize}
The graph $T_p(Q)$ has $2n_Q+4p(m_Q+n_Q)$ vertices and can be constructed in time polynomial in $n_Q$ and $p$. Every edge of $T_p(Q)$ either joins two core vertices or is incident with exactly one attachment, in the sense that it is incident with the middle vertex or with a private vertex of that attachment. Index $i$ refers to the two copies of $Q$, and index $j$ to the $p$ repetitions of the attachments.

\begin{lemma}\label{lem:chordal-structure}
The graph $T_p(Q)$ is chordal, and a perfect elimination ordering of it can be computed in polynomial time. Deleting its private vertices leaves a split graph, and every private vertex has degree two, and its two neighbors are adjacent.
\end{lemma}

\begin{proof}
Order the vertices with all private vertices first, then all middle vertices, then all core vertices. The later neighbors of a private vertex are its two neighbors $c^i_v$ and $a^{i,j}_v$, which are adjacent. The later neighbors of a middle vertex are core vertices, which form a clique, and so do the later neighbors of a core vertex. Hence this ordering, which is computed along with $T_p(Q)$, is a perfect elimination ordering, and $T_p(Q)$ is chordal. After deleting the private vertices, the core vertices form a clique and the middle vertices an independent set, so the remaining graph is a split graph.
\end{proof}

For a set $S$ of vertices of $T_p(Q)$, let $S^i=\{v\in V(Q):c^i_v\in S\}$ for $i\in\{1,2\}$, and let $\rho^i_v$ be the number of neighbors $u$ of $v$ in $Q$ such that exactly one of $c^i_u$ and $c^i_v$ lies in $S$. Every edge of $Q$ that is cut by $S^i$ is counted at both of its endpoints, so
\begin{equation}\label{eq:rho-sum}
\sum_{v\in V(Q)}\rho^i_v=2\cut_Q(S^i).
\end{equation}

\begin{lemma}\label{lem:local}
For every set $S$ of vertices of $T_p(Q)$, the number of edges of the attachment $(i,v,j)$ that are cut by $S$ is at most $2(\deg_Q(v)+1)+\rho^i_v$. Equality holds when $a^{i,j}_v$ lies on the same side as $c^i_v$ and all private vertices of the attachment lie on the other side.
\end{lemma}

\begin{proof}
Write $d=\deg_Q(v)$. Suppose first that $a^{i,j}_v$ lies on the same side as $c^i_v$. Then the edge $a^{i,j}_vc^i_v$ is not cut, exactly $\rho^i_v$ of the edges from $a^{i,j}_v$ to the vertices $c^i_u$ are cut, and each private vertex has both neighbors on one side, so at most two of its edges are cut, and exactly two if it lies on the other side. The total is at most $2(d+1)+\rho^i_v$, with the stated case of equality. Suppose now that $a^{i,j}_v$ lies on the other side than $c^i_v$. Then the edge $a^{i,j}_vc^i_v$ is cut, exactly $d-\rho^i_v$ of the edges to the vertices $c^i_u$ are cut, and each private vertex has its two neighbors on different sides, so exactly one of its edges is cut. The total is $1+(d-\rho^i_v)+(d+1)=2(d+1)-\rho^i_v$, which is at most $2(d+1)+\rho^i_v$.
\end{proof}

When $\rho^i_v=0$ the two placements of the middle vertex are equally good; the lemma only asserts that the first one is optimal, which is all we need. The number $\deg_Q(v)+1$ of private vertices is the smallest for which the first placement is optimal for every value of $\rho^i_v$: with $q$ private vertices, the best placements of the two kinds cut $2q+\rho^i_v$ and $q+1+\deg_Q(v)-\rho^i_v$ edges of the attachment. In particular, without private vertices the best placement of the middle vertex would cut $\max\{\rho^i_v,\deg_Q(v)+1-\rho^i_v\}$ edges of the attachment, which is not an increasing function of $\rho^i_v$. The private vertices are thus essential, and deleting them leaves a split graph (\cref{lem:chordal-structure}); this is why the reduction does not give hardness on split graphs themselves. A similar use of degree-two vertices adjacent to both endpoints of an edge, which favor placing these endpoints on the same side, appears in the reduction of Bodlaender and Jansen for chordal graphs~\cite{bj2000}; here the number of private vertices is calibrated to the degree in $Q$, and the two complementary copies and the amplification give the quantitative relation of \cref{lem:decoder}.

\begin{lemma}\label{lem:decoder}
Let $o=\mc(Q)$. Then
\[
\mc(T_p(Q))=4p\,(2m_Q+n_Q+o)+n_Q^2 .
\]
Moreover, every set $S$ of vertices of $T_p(Q)$ satisfies
\[
\max\bigl\{\cut_Q(S^1),\cut_Q(S^2)\bigr\}\ \ge\ o-\frac{\mc(T_p(Q))-\cut_{T_p(Q)}(S)}{4p}.
\]
\end{lemma}

\begin{proof}
Let $S$ be a set of vertices of $T_p(Q)$, let $\kappa$ be the number of core vertices in $S$, and let $o'=\max\{\cut_Q(S^1),\cut_Q(S^2)\}$. Since every edge of $T_p(Q)$ joins two core vertices or belongs to exactly one attachment,
\[
\cut_{T_p(Q)}(S)=\kappa(2n_Q-\kappa)+\sum_{i\in\{1,2\}}\ \sum_{v\in V(Q)}\ \sum_{j=1}^{p}e_{i,v,j}(S),
\]
where $e_{i,v,j}(S)$ is the number of edges of the attachment $(i,v,j)$ that are cut by $S$. The first term is at most $n_Q^2$. By \cref{lem:local} and~\eqref{eq:rho-sum}, the triple sum is at most
\[
\sum_{i\in\{1,2\}}\ \sum_{v\in V(Q)}p\bigl(2(\deg_Q(v)+1)+\rho^i_v\bigr)=\sum_{i\in\{1,2\}}p\bigl(4m_Q+2n_Q+2\cut_Q(S^i)\bigr)\le4p\,(2m_Q+n_Q+o') .
\]
Hence $\cut_{T_p(Q)}(S)\le4p\,(2m_Q+n_Q+o')+n_Q^2$. Since $o'\le o$, this shows $\mc(T_p(Q))\le4p\,(2m_Q+n_Q+o)+n_Q^2$, and it also gives the second claim once the first is proved.

For the matching lower bound, let $O$ be a maximum cut of $Q$, and let $S$ contain $c^1_v$ for $v\in O$ and $c^2_v$ for $v\notin O$. Then $S$ contains exactly $n_Q$ core vertices, so the core contributes $n_Q^2$ cut edges, and $\cut_Q(S^1)=\cut_Q(O)=o$ and $\cut_Q(S^2)=\cut_Q(V(Q)\setminus O)=o$. Placing the middle and private vertices as in the case of equality of \cref{lem:local} makes every attachment contribute $2(\deg_Q(v)+1)+\rho^i_v$, so the attachments contribute $\sum_{i}p(4m_Q+2n_Q+2o)=4p\,(2m_Q+n_Q+o)$, and $\cut_{T_p(Q)}(S)=4p\,(2m_Q+n_Q+o)+n_Q^2$.
\end{proof}

\subsection{Proof of the hardness theorem}\label{sec:hardness-proof}

\begin{restatedtheorem}{thm:hardness}
\thmHardnessText
\end{restatedtheorem}

\begin{proof}
An algorithm that approximates Max-Cut within a factor $\beta'$ also approximates it within every factor $\beta\le\beta'$. It therefore suffices to prove the theorem for $0<\eps<1/85$. Fix such an $\eps$, let $\beta=84/85+\eps$, and suppose that a polynomial-time algorithm $\mathcal{B}$ approximates Max-Cut within $\beta$ on graphs obtained from a split graph by adding vertices of degree two whose two neighbors are adjacent. We derive a polynomial-time algorithm that approximates Max-Cut within $16/17+4\eps$ on simple graphs, which implies $\Pclass=\NP$ by \cref{thm:source}.

Let $Q_0$ be a simple graph with $n_0$ vertices and $m_0\ge1$ edges; graphs without edges are trivial. Let
\[
k=\Bigl\lceil\frac{4n_0}{\eps m_0}\Bigr\rceil,\qquad p=\Bigl\lceil\frac{n_0^2}{\eps m_0}\Bigr\rceil,\qquad Q=Q_0^{(k)} .
\]
Then $n_Q=kn_0$ and $m_Q=k^2m_0$, and $k$ and $p$ are polynomial in $n_0$ for fixed $\eps$. We construct $T=T_p(Q)$, which belongs to the class by \cref{lem:chordal-structure}, and run $\mathcal{B}$ on it to obtain a set $S$ with $\cut_T(S)\ge\beta\mc(T)$. Let $o'=\max\{\cut_Q(S^1),\cut_Q(S^2)\}$ and $o=\mc(Q)$. By \cref{lem:decoder},
\[
o-o'\le\frac{(1-\beta)\mc(T)}{4p}=(1-\beta)\Bigl(2m_Q+n_Q+o+\frac{n_Q^2}{4p}\Bigr).
\]
Since $o\ge m_Q/2$ by~\eqref{eq:half}, we have $2m_Q\le4o$, $n_Q\le(2n_Q/m_Q)\,o$, and $n_Q^2/(4p)\le(n_Q^2/(2pm_Q))\,o$. By the choice of $k$ and $p$,
\[
\frac{2n_Q}{m_Q}=\frac{2n_0}{km_0}\le\frac\eps2\qquad\text{and}\qquad\frac{n_Q^2}{2pm_Q}=\frac{n_0^2}{2pm_0}\le\frac\eps2 .
\]
Hence $o'\ge\bigl(1-(5+\eps)(1-\beta)\bigr)o$. Applying \cref{lem:blowup} to the better of $S^1$ and $S^2$ gives a set $S_0$ with
\[
\cut_{Q_0}(S_0)\ge\frac{o'}{k^2}\ge\bigl(1-(5+\eps)(1-\beta)\bigr)\frac{o}{k^2}=\bigl(1-(5+\eps)(1-\beta)\bigr)\mc(Q_0).
\]
Finally, since $1-\beta=1/85-\eps$,
\[
1-(5+\eps)\Bigl(\frac1{85}-\eps\Bigr)=\frac{16}{17}+5\eps-\frac{\eps}{85}+\eps^2>\frac{16}{17}+4\eps .
\]
All steps run in polynomial time, so we have obtained a polynomial-time algorithm that approximates Max-Cut on simple graphs within $16/17+4\eps$.
\end{proof}

\section*{AI methodology}
\addcontentsline{toc}{section}{AI methodology}
Generative AI tools, including ChatGPT and Claude, were used extensively throughout this work.  They
contributed to the formulation of theorems, algorithms, proofs, and other
claims.  They were also used for literature search and citation preparation,
figure production, and the drafting and iterative
revision of the manuscript.  The authors guided and reviewed this process.
The authors assume full responsibility for the content.

\IfFileExists{alphaurl.bst}{\bibliographystyle{alphaurl}}{\bibliographystyle{alpha}}
\phantomsection\addcontentsline{toc}{section}{References}
\bibliography{interval_maxcut_arxiv_references_v1}

\end{document}